\documentstyle[11pt]{article}
\catcode`@=11
\def\seceqaa{\@addtoreset{equation}{section}
           \def\theequation{A\arabic{equation}}}
\def\seceqbb{\@addtoreset{equation}{section}
           \def\theequation{B\arabic{equation}}}
\def\seceqcc{\@addtoreset{equation}{section}
           \def\theequation{C\arabic{equation}}}
\def\seceqdd{\@addtoreset{equation}{section}
           \def\theequation{C\arabic{equation}}}
\catcode`@=11
\begin{document}
\begin{titlepage}
\begin{center}
{\Large \bf Moduli Stabilization, Large-Volume dS Minimum Without $\overline{D3}$-Branes, (Non-)Supersymmetric
Black Hole Attractors and Two-Parameter Swiss Cheese Calabi-Yau's}
\vskip 0.1in { Aalok Misra$^{(a),(b)}$\footnote{e-mail: aalokfph@iitr.ernet.in
} and
Pramod Shukla$^{(a)}$ \footnote{email: pmathdph@iitr.ernet.in}\\
(a) Department of Physics, Indian Institute of Technology,
Roorkee - 247 667, Uttaranchal, India\\
(b) Physics Department, Theory Unit, CERN, CH1211, Geneva 23, Switzerland } \vskip 0.5 true in
\end{center}
\thispagestyle{empty}
\begin{abstract}
We consider two sets of issues in this paper. The first has to do with moduli stabilization, existence of ``area codes" \cite{Giryavets} and the possibility of getting a non-supersymmetric dS minimum without the addition of $\overline{D3}$-branes as in KKLT
for type II flux compactifications. The second has to do with
the ``Inverse Problem" \cite{VafaInverse} and ``Fake Superpotentials" \cite{Ceresole+Dall'agata} for extremal (non)supersymmetric black holes in
type II compactifications. We use (orientifold of) a ``Swiss Cheese" Calabi-Yau \cite{SwissCheese} expressed as a degree-18 hypersurface in ${\bf WCP}^4[1,1,1,6,9]$ in the ``large-volume-scenario" limit \cite{Balaetal2}. The main result of our paper is that we show that by including non-perturbative $\alpha^\prime$ and instanton corrections in the
K\"{a}hler potential and superpotential \cite{Grimm}, it may be possible to obtain a large-volume non-supersymmetric
dS minimum {\it without} the addition of anti-D3 branes a la KKLT. The chosen Calabi-Yau has
been of relevance also from the point of other studies of K\"{a}hler moduli stabilization via
nonperturbative instanton contributions \cite{Denef+Douglas+Florea} and non-supersymmetric AdS vacua (and their subsequent dS-uplifts) using $(\alpha^\prime)^3$ corrections to the K\"{a}hler potential \cite{Balaetal2,BBHL,westphal,Balaetal1}.
\end{abstract}
\end{titlepage}
\newpage
\section{Introduction}

Flux compactifications have been extensively studied from the point of view of moduli stabilization
(See \cite{Grana} and references therein). Though, generically only the complex structure moduli get stabilized
by turning on fluxes and one needs to consider non-perturbative moduli stabilization for the K\"{a}hler
moduli\cite{KKLT}. In the context of type II compactifications, it is naturally interesting to look for examples
wherein it may be possible to stabilize the complex structure moduli (and the axion-dilaton modulus)
at different points of the moduli space that are finitely separated, for the {\it same} value of the fluxes. This phenomenon is referred to as ``area codes" that leads to formation of domain walls.
Another extremely important issue related to moduli stabilization is the problem of getting a non-supersymmetric de Sitter vacuum in string theory. The KKLT scenario which even though does precisely that, has the problem of addition of an uplift term to the the potential, corresponding to addition of $\overline{D3}$-branes, that can not be cast into an ${\cal N}=1$ SUGRA formalism. It would be interesting to be able to get a de Sitter vacuum without the addition of such $\overline{D3}$-branes. The Large Volume Scenarios' study initiated in \cite{Balaetal2} provides a hope for the same. Further,  there is a close connection between flux vacua and black-hole attractors. It has been shown that extremal black holes exhibit an interesting phenomenon - the attractor mechanism \cite{attractor}. In the same, the moduli are ``attracted" to some fixed values determined by the charges of the black hole, independent of the asymptotic values of the moduli. Supersymmetric black holes at the attractor point, correspond to
minimizing the central charge and the effective black hole potential, whereas nonsupersymmetric attractors \cite{nonsusybh1}, at the
attractor point, correspond to minimizing only the potential and not the central charge. The latter have recently been (re)discussed \cite{nonsusybh2} in the literature.

In this paper, we try to address all the issues of the previous paragraph by exploring different perturbative and non-perturbative (in $\alpha^\prime$ and instanton contributions) aspects of (non)supersymmetric flux vacua and
black holes in the context of type II compactifications on (orientifold) of compact Calabi-Yau's of a
projective variety with multiple singular conifold loci in their moduli space. The compact Calabi-Yau we work with is of the ``Swiss cheese" type. The paper is planned as follows. In section {\bf 2}, based on \cite{Candelasetal}, we perform a detailed analysis of the periods of the Calabi-Yau three-fold considered in this paper, working out their forms in the symplectic basis for points away and close to the two singular conifold loci. The  results of section {\bf 2} get used in the subsequent section ({\bf 3}). We then discuss, in section {\bf 3}, stabilization of the
complex structure moduli including the axion-dilaton modulus by extremizing the flux superpotential for points
near and close to the two conifold loci, arguing the existence of ``area codes" and domain walls.
In section {\bf 4}, we show that by the inclusion of non-perturbative $\alpha^\prime$-corrections to
the K\"{a}hler potential that survive orientifolding and instanton contributions to the superpotential, one
can, analogous to \cite{Balaetal2}, get a large-volume non-supersymmetric dS vacuum {\it without the addition of $\overline{D3}$-branes}. We consider this to be the most significant result of this paper.
In section {\bf 5}, we explicitly solve the ``inverse problem" using the techniques of \cite{VafaInverse}.
In section {\bf 6}, using the techniques of \cite{Ceresole+Dall'agata} we show the existence of multiple superpotentials (including therefore ``fake superpotentials").
Section {\bf 7} has the conclusions.

\section{The Moduli Space Scan and the Periods}

In this section, based on results in \cite{Candelasetal}, we look at different regions in the moduli space of a two-parameter Calabi-Yau three fold of a
projective variety expressed as a hypersurface in a weighted complex projective space, and write out the explicit
expressions for the periods. The explicit expressions, though cumbersome, will be extremely useful when studying complex structure moduli stabilization and existence of ``area codes" in section {\bf 3}, solving explicitly the ``inverse problem" in section {\bf 4} and showing explicitly the existence of ``fake superpotentials" in section {\bf 5} in the context of non-supersymmetric black hole attractors. More precisely,
based on \cite{Candelasetal}, we will consider the periods of the ``Swiss cheese" \footnote{The term ``Swiss cheese" (See \cite{SwissCheese}) is used to denote those Calabi-Yau's whose volume can be written as: ${\cal V}=(\tau^B + \sum_{i\neq B} a_i\tau^S_i)^{\frac{3}{2}} - (\sum_{j\neq B}b_j\tau^S_j)^{\frac{3}{2}} - ...$, where $\tau^B$ is the volume of the big divisor and $\tau^S_i$ are the volumes of the $h^{1,1}-1$ (corresponding to the (1,$h^{1,1}-1$)-signature of the Hessian) small divisors. The big divisor governs the size of the Swiss cheese and the small divisors control the size of the holes of the same Swiss cheese.} Calabi-Yau three-fold obtained as a resolution of the degree-18 hypersurface in ${\bf WCP}^4[1,1,1,6,9]$:
\begin{equation}
\label{eq:hypersurface}
x_1^{18} + x_2^{18} + x_3^{18} + x_4^3 + x_5^2 - 18\psi \prod_{i=1}^5x_i - 3\phi x_1^6x_2^6x_3^6 = 0.
\end{equation}
Similar to the explanation given in \cite{Kachruetal}, it is understood that only two complex structure
moduli $\psi$ and $\phi$ are retained in (\ref{eq:hypersurface}) which are invariant under the group $G$ of
footnote 3, setting the other invariant complex structure moduli appearing at a higher order (due to invariance under $G$)
 at their values at the origin.

Defining $\rho\equiv (3^4.2)^{\frac{1}{3}}\psi$, the singular loci of (\ref{eq:hypersurface})
are in ${\bf WCP}^2[3,1,1]$ with homogenous coordinates $[1,\rho^6,\phi]$ and are given as under:
\begin{enumerate}
\item
${\it Conifold\ Locus 1}: \{(\rho,\phi)|(\rho^6+\phi)^3=1\}$
\item
${\it Conifold\ Locus 2}: \{(\rho,\phi)|\phi^3=1\}$
\item
${\it Boundary}: (\rho,\phi)\rightarrow\infty$
\item
${\it Fixed\ point\ of\ quotienting}$: The fixed point $\rho=0$ of ${\cal A}^3$ where
${\cal A}:(\rho,\phi)\rightarrow(\alpha\rho,\alpha^6\phi)$, where $\alpha\equiv e^{\frac{2\pi i}{18}}
$\footnote{This is induced by the group action:
$(x_1,x_2,x_3,x_4,x_5;\psi,\phi)\rightarrow(\alpha^{A_1}x_1,\alpha^{A_2}x_2,\alpha^{A_3}x_3,
\alpha^{6A_4}x_4,\alpha^{9A_5}x_5;
\alpha^{-a}\psi,\alpha^{-6a}\phi)$ where $a=A_1 + A_2 + A_3 + 6A_4 + 9A_5$ and
$(A_1,A_2,A_3,A_4,A_5)$ are related to the coefficients of the most general degree-18 polynomial
in $(x_1,x_2,x_3,x_4,x_5)$ invariant under
$G={\bf Z}_6\times{\bf Z}_{18}$ (${\bf Z}_6:(0,1,3,2,0,0); {\bf Z}_{18}:(1,-1,0,0,0)$).  The mirror to
$p=x_1^{18} + x_2^{18} + x_3^{18} + x_4^3 + x_5^2 =0$, according to the Greene-Plesser construction, is given
by $\{p=0\}/G$. The $G-$invariant polynomial is given by:
$A_0^\prime\prod_{i=1}^6x_i + A_1^\prime x_1^2 x_2^2 x_3^2 x_4^2
+ A_2^\prime x_1^3 x_2^3 x_3^3 x_5 + A_3^\prime x_1^4 x_2^4 x_3^4 x_4 + A_4^\prime x_1^6 x_2^6x_3^6
+ A_5^\prime x_1^{18} + A_6^\prime x_2^{18} + A_7^\prime x_3^{18} + A_8^\prime x_4^3 + A_9^\prime x_5^2$;
some of the deformations can be redefined away by suitable automorphisms.}.
\end{enumerate}

We will be considering the following sectors in the $(\rho,\phi)$ moduli space:
\begin{itemize}
\item
$\underline{|\phi^3|>1, 0<arg\phi<\frac{2\pi}{3},{\rm large}\ \psi}$

The fundamental period $\varpi_0$, obtained by directly integrating the holomorphic three-form over the ``fundamental cycle" (See \cite{Candelasetal}), is given by:
\begin{eqnarray}
\label{eq:largephilargepsi1}
& & \varpi_0=\sum_{k=0}^\infty \frac{(6k)!}{k!(2k)!(3k)!}\left(\frac{-3}{18^6\psi^6}\right) U_k(\phi)\nonumber\\
& & =\sum_{k=0}^\infty\frac{(-)^k\Gamma(k+\frac{1}{6})\Gamma(k+\frac{5}{6})}{(k!)^2}\left(\frac{1}{\rho^{6k}}\right)
U_k(\phi),
\end{eqnarray}
where $U_\nu(\phi)\equiv \phi^\nu\ _3F_2(-\frac{\nu}{3},\frac{1-\nu}{3},\frac{2-\nu}{3};1,1;\frac{1}{\phi^3})$; the other components of the period vector are given by: $\varpi_i=\varpi_0(\alpha^i\psi,\alpha^{6i}\phi)$ where
$\alpha\equiv e^{\frac{2\pi i}{18}},\ i=1,2,3,4,5$.

\item
$\underline{|\phi^3|<1,\ {\rm large}\ \psi}$

The fundamental period is given by:
\begin{equation}
\label{eq:smallphilargepsi11}
\varpi_0=\sum_{m=0}^\infty\sum_{n=0}^\infty\frac{(18n + 6m)!(-3\phi)^m}{(9n + 3m)!(6n + 2m)!(n!)^3m!(18\psi)^{18n + 6m}},
\end{equation}
implying that around a suitable $\rho=\rho_0$ and $\phi=\phi_0$:
\begin{equation}
\label{eq:smallphilargepsi2}
\left(\begin{array}{c}
\varpi_0\\
\varpi_1\\
\varpi_2\\
\varpi_3\\
\varpi_4\\
\varpi_5\end{array}\right)
=\left(\begin{array}{ccc}P_1 & P_2 & P_3\end{array}\right)\left(\begin{array}{c} 1\\(\phi - \phi_0)\\(\rho - \rho_0)
\end{array}\right),
\end{equation}
where $P_{1,2,3}$ are given in appendix A.

\item
$\underline{|\frac{\rho^6}{\phi - \omega^{0,-1,-2}}|<1}$

\begin{equation}
\label{eq:awaycl11}
\varpi_{3a+\sigma}=\frac{1}{3\pi}\sum_{r=1,5}\alpha^{3ar}sin\left(\frac{\pi r}{3}\right)\xi^\sigma_r(\psi,\phi)[a=0,1;\sigma=0,1,2],
\end{equation}
where $\xi_r^\sigma(\psi,\phi)=\sum_{k=0}^\infty\frac{(\Gamma(k+\frac{r}{6})^2}{k!\Gamma(k+\frac{r}{3})}\rho^{6k+r}
U_{-(k+\frac{r}{6})}^\sigma(\phi), U^\sigma_\nu(\phi)=\omega^{-\nu\sigma}U_\nu(\omega^\sigma\phi)\ \footnote{The three values of $\sigma$ correspond to the three solutions to $(1-\phi^3)U_\nu^{\prime\prime\prime}(\phi)
+3(\nu-1)\phi^2U^{\prime\prime}_\nu(\phi) - (3\nu^2-3\nu+1)\phi U^\prime_\nu(\phi)+\nu^3 U_\nu(\phi)=0$; the
Wronskian of the three solutions is given by: $\frac{-27i}{2\pi^3}e^{-i\pi\nu}sin^2(\pi\nu)(1-\phi^3)^{\nu-1}$ -
the solutions are hence linearly independent except when $\nu\in{\bf Z}$}, \omega\equiv e^{\frac{2\pi i}{3}}$; for small $\phi$,
\begin{equation}
\label{eq:Usmallphi}
U_\nu(\phi)=\frac{3^{-1-\nu}}{\Gamma(-\nu)}\sum_{m=0}^\infty\frac{\Gamma(\frac{m-\nu}{3})(3\omega\phi)^m}{(\Gamma(1 - \frac{m-\nu}{3})^2m!}.
\end{equation}
Expanding about a suitable $\phi=\phi_0$ and $\rho=\rho_0$, one can show:
\begin{equation}
\label{eq:awaycl12}
\left(\begin{array}{c}
\varpi_0\\
\varpi_1\\
\varpi_2\\
\varpi_3\\
\varpi_4\\
\varpi_5\\
\end{array}\right)=\left(\begin{array}{ccc}
M_1 & M_2 & M_3
\end{array}\right)\left(\begin{array}{c}
1\\
(\rho - \rho_0)\\
(\phi - \phi_0)\\
\end{array}\right),
\end{equation}
where
$M_{1,2,3}$ are given in appendix A.

\item
$\underline{{\rm Near\ the\ conifold\ locus:}\ \rho^6 + \phi = 1}$

The periods are given by:
\begin{equation}
\label{eq:confl11}
\varpi_i = C_i g(\rho,\phi) ln(\rho^6 + \phi - 1) + f_i(\rho,\phi),
\end{equation}
where $C_i=(1,1,-2,1,0,0)$, $g(\rho,\phi)=\frac{i}{2\pi}(\varpi_1 - \varpi_0)\sim a(\rho^6 + \phi - 1)$ near $\rho^6 + \phi - 1\sim0$ where $a$ is a constant and $f_i$ are analytic in $\rho$ and $\psi$. The
analytic functions near the conifold locus are given by:
\begin{equation}
\label{eq:confl12}
f_{3a + \sigma}=\frac{1}{2\pi}\sum_{r=1,5}e^{\frac{i\pi ar}{3}}sin\left(\frac{\pi r}{3}\right) \xi_r^\sigma(\rho,\phi), a=0,1;\ \sigma=0,1,2.
\end{equation}
Defining $x\equiv(\rho^6 + \phi - 1)$, one can show that:
\begin{equation}
\label{eq:confl13}
\xi_r^\sigma=\sum_{k=0}^\infty\sum_{m=0}^\infty
\frac{3^{-1 + k + \frac{r}{6} + m}e^{\frac{2i\pi(\sigma + 1)}{3} + \frac{-i\pi(k + \frac{r}{6})}{3}}(-)^k(\Gamma(k + \frac{r}{6}))^2}{\Gamma(k + 1)\Gamma(k + \frac{r}{6})\Gamma(k + \frac{r}{6})(\Gamma(1 - \frac{m + k + \frac{r}{6}}{3}))^2m!}(x - \phi + 1)^{k + \frac{r}{6}}.
\end{equation}
One can hence see that:
\begin{equation}
\label{eq:confl14}
\left(\begin{array}{c}
f_0\\
f_1\\
f_2\\
f_3\\
f_4\\
f_5\end{array}\right)=\left(\begin{array}{ccc} N_1 & N_2 & N_3\end{array}\right)\left(\begin{array}{c}
1\\
x\\
\phi\end{array}\right),
\end{equation}
where $N_{1,2,3}$ are given in appendix A.

\item
$\underline{{\rm Near}\ \phi^3=1,\ {\rm Large}\ \rho}$

From asymptotic analysis of the coefficients, one can argue:
\begin{eqnarray}
\label{eq:confl21}
& & U_\nu(\phi)\sim-\frac{\sqrt{3}}{2\pi(\nu+1)}\left[(\phi-1)^{\nu+1} - 2\omega(\phi-\omega^{-1}) + \omega^2(\phi-\omega^{-2})\right],\nonumber\\
& & \equiv y^0_\nu - 2y^1_\nu + y^2_\nu,
\end{eqnarray}
where $\omega\equiv e^{\frac{2i\pi}{3}}$. Defining $U^\sigma_\nu(\phi)=\sum_{\tau=0}^2\gamma_\nu^{\sigma,\tau}y_\nu^\tau(\phi)$, where
$\gamma^{\sigma,\tau}_\nu=\left(\begin{array}{ccc}
1 & -2 & 1 \\
e^{-2i\pi\nu} & 1 & -2 \\
-2e^{-2i\pi\nu} & e^{-2i\pi\nu} & 1\\
\end{array}\right)$, one can show that $U_\nu,\left(\frac{\sum_{\sigma=0}^2U^\sigma_\nu(\phi)}{1-e^{-2i\pi\nu}}=\right)y^0_\nu(\phi) - y^1_\nu(\phi)\equiv V_\nu(\phi),\left(\frac{3V_\nu(\phi) - 2U_\nu(\phi) - U^1_\nu(\phi)}{1-e^{-2i\pi\nu}}=\right)y^0_\nu(\phi)\equiv W_\nu(\phi)$ are linearly independent even for $\nu\in{\bf Z}$\footnote{The Wronskian of these three solutions
is given by $\frac{27i}{(2\pi)^3}e^{i\pi\nu}(1-\phi^3)^{\nu-1}\neq0,\nu\in{\bf Z}$.}.

For small $\rho$,
\begin{equation}
\label{eq:confl22}
\xi^\sigma_r=\int_\Gamma \frac{d\mu}{2i sin(\pi(\mu+\frac{r}{6}))}
\frac{(\Gamma(-\mu))^2}{\Gamma(-\mu+\frac{1}{6})\Gamma(-\mu+\frac{5}{6})}\rho^{-6\mu}U^\sigma_\mu(\phi),
\end{equation}
where the contour $\Gamma$ goes around the Im$(\mu)<0$ axis. To deform the contour to a contour $\Gamma^\prime$ going around the Im$(\mu)>0$ axis, one sees that one can do so for $\sigma=0$ but not for $\sigma=1,2$. For the latter, one modifies $U^\sigma_\mu(\phi)$ by adding a function which does not contribute to the poles and has simple zeros at integers as follows:
\begin{equation}
\label{eq:confl23}
U^\sigma_\mu(\phi)\rightarrow\tilde{U^\sigma_{\mu,r}(\phi)}\equiv U^\sigma_\mu(\phi) - e^{\frac{i\pi r}{6}}\frac{sin(\pi(\mu+\frac{r}{6}))}{sin(\pi\mu)}f^\sigma_\mu(\phi),
\end{equation}
where
\begin{eqnarray}
\label{eq:confl231}
& & f^0_\mu(\phi)=0,\nonumber\\
& & f^1_\mu(\phi)=-(1-e^{-2i\pi\nu})y^0_\mu(\phi),\nonumber\\
& & f^2_\mu(\phi)=(1-e^{-2i\pi\nu})V_\nu(\phi) + (1-e^{-2i\pi\nu})W_\mu(\phi).
\end{eqnarray}
One can then deform the contour $\Gamma$ to the contour $\Gamma^\prime$ to evaluate the periods. This is done in appendix A. 

Expanding about $\phi=\omega^{-1}$, and a large $\rho=\rho_0$, one gets the following periods:
\begin{equation}
\label{eq:confl210}
\left(\begin{array}{c}
\varpi_0\\
\varpi_1\\
\varpi_2\\
\varpi_3\\
\varpi_4\\
\varpi_5\\
\end{array}\right)=\left(\begin{array}{c}
A^\prime_0 + B^\prime_{01}x + C^\prime_0(\rho - \rho_0)\\
A^\prime_1 + B^\prime_{11}x + C^\prime_1(\rho - \rho_0)\\
A^\prime_2 + B^\prime_{21}x + C^\prime_2(\rho - \rho_0)\\
A^\prime_3 + B^\prime_{31}x + B^\prime_{32}\ x\ ln x + C^\prime_3(\rho - \rho_0)\\
A^\prime_4 + B^\prime_{41}x + B^\prime_{42}\ x\ ln x + C^\prime_4(\rho - \rho_0)\\
A^\prime_5 + B^\prime_{51}x + B^\prime_{52}\ x\ ln x + C^\prime_5(\rho - \rho_0)\\
\end{array}\right),
\end{equation}
where $x\equiv(\phi - \omega^{-1})$.
\end{itemize}
The equations (\ref{eq:confl24}) and (\ref{eq:confl210}) will get used to arrive at (\ref{eq:Wconfl11}) and finally
(\ref{eq:Wconfl161}) and (\ref{eq:Wconfl17}).

The Picard-Fuchs basis of periods evaluated above can be transformed to a symplectic basis as under
(See \cite{Candelasetal}):
\begin{equation}
\label{eq:PFbasis1}
\Pi=\left(\begin{array}{c}
F_0\\
F_1\\
F_2\\
X^0\\
X^1\\
X^2
\end{array}\right) = M \left(\begin{array}{c}
\varpi_0\\
\varpi_1\\
\varpi_2\\
\varpi_3\\
\varpi_4\\
\varpi_4
\end{array}\right),
\end{equation}
where
\begin{equation}
\label{eq:PFbasis2}
M=\left(\begin{array}{cccccc}
-1&1&0&0&0&0\\
1&3&3&2&1&0\\
0&1&1&1&0&0\\
1&0&0&0&0&0\\
-1&0&0&1&0&0\\
2&0&0&-2&1&1
\end{array}\right).
\end{equation}
In the next section, we use information about the periods evaluated in this section, in looking for ``area codes".

\section{Extremization of the Superpotential and Existence of ``Area Codes"}

In this section, we argue the existence of area codes, i.e., points in the moduli space close to and away from the two singular conifold loci that are finitely
separated where for the same large values (and hence not necessarily integral) of RR and NS-NS fluxes, one
can extremize the (complex structure and axion-dilaton) superpotential (for different values of the complex structure and axion-dilaton moduli)\footnote{
For techniques in special geometry relevant to this work, see \cite{Mohaupt} for a recent review; see
\cite{Curio+Spillner} for moduli-stablization calculations as well.}.

The axion-dilaton modulus $\tau$ gets stabilized (from $D_\tau W_{c.s.}=0$, $W_{c.s.}$ being the Gukov-Vafa-Witten complex structure superpotential $\int (F_3 - \tau H_3)\wedge\Omega=(2\pi)^2\alpha^\prime(f-\tau h)\cdot\Pi$, $F_3$ and $H_3$ being respectively the NS-NS and RR three-form field strengths, and are given by:
$F_3=(2\pi)^2\alpha^\prime\sum_{a=1}^3(f_a\beta_a + f_{a+3}\alpha_a)$ and
$H_3=(2\pi)^2\alpha^\prime\sum_{a=1}^3(h_a\beta_a+h_{a+3}\alpha_a)$;
$\alpha_a,\beta^a$,  $a=1,2,3$,
form an integral cohomology basis) at a value given by:
\begin{equation}
\label{eq:taufix1}
\tau=\frac{f^T.{\bar\Pi_0}}{h^T.{\bar\Pi_0}},
\end{equation}
where $f$ and $h$ are the fluxes corresponding to the NS-NS and RR fluxes; it is understood that the complex structure moduli appearing in (\ref{eq:taufix1}) are already fixed from $D_iW=0,\ i=1,2$.

\begin{itemize}
\item
$\underline{{\rm Near\ the\ conifold\ locus:}\ \phi^3=1,\ {\rm Large}\ \psi}$

The period vector in the symplectic basis, is given by:
\begin{eqnarray}
\label{eq:Wconfl11}
& & \Pi=\nonumber\\
& & \hskip - 2.8cm\left(\begin{array}{c}
-A^\prime_0 + A^\prime_1 + (C^\prime_1 - C^\prime_0) (\rho - \rho_0) + (B^\prime_{11} - B^\prime_{01}) x \\
(A^\prime_0 + A^\prime_4 + 3 A^\prime_1 + 3 A^\prime_2 + 2A^\prime_3) + (C^\prime_0 + C^\prime_4) (\rho - \rho_0) + (B^\prime_{01} + B^\prime_{41} + 3 B^\prime_{11} + 3B^\prime_{21} + 2 B^\prime_{31})x + (3 B^\prime_{42} + 2 B^\prime_{32}) x ln x \\
A^\prime_1 + A^\prime_2 + A^\prime_3 + (B^\prime_{11} + B^\prime_{21} + B^\prime_{31}) x + (C^\prime_1 + C^\prime_2 + C^\prime_3) (\rho - \rho_0) + B^\prime_{32} x ln x \\
A^\prime_0 + C^\prime_{01} x + C^\prime_0 (\rho - \rho_0) \\
- A^\prime_0 + C^\prime_{01} x + C^\prime_0 (\rho - \rho_0) \\
- A^\prime_0 + A^\prime_3 + (B^\prime_{31} - B^\prime_{01})x + (C^\prime_3 - C^\prime_0) (\rho - \rho_0) + B^\prime_{32} x ln x\\
2 A^\prime_0 - 2 A^\prime_3 + A^\prime_4 + A^\prime_5 + (2 B^\prime_{01} - 2 B^\prime_{31} + B^\prime_{41} + B^\prime_{51})x  + (2C^\prime_0 - 2 C^\prime_3 + C^\prime_4 + C^\prime_5) (\rho - \rho_0) + (-2 B^\prime_{32} + B^\prime_{42} + B^\prime_{52}) x ln x
\end{array}\right)\nonumber\\
& & \equiv \left(\begin{array}{c} A_0 + B_{01} x + C_0 (\rho - \rho_0) \\ A_1 + B_{11} x + B_{12} x ln x + C_1 (\rho - \rho_0) \\ A_2 + B_{21} x + B_{22} x ln x + C_2 (\rho - \rho_0) \\ A_3 + B_{31} x + C_3 (\rho - \rho_0) \\ A_4 + B_{41} x + B_{42} x ln x + C_4 (\rho - \rho_0)\\ A_5 + B_{51} x + B_{52} x ln x + C_5 (\rho - \rho_0) \end{array} \right).
\end{eqnarray}
The tree-level K\"{a}hler potential is given by:
\begin{equation}
\label{eq:Wconfl12}
K = -ln\left(-i(\tau - {\bar\tau})\right) - ln\left(-i\Pi^\dagger\Sigma\Pi\right),
\end{equation}
where the symplectic metric $\Sigma=\left(\begin{array}{cc} 0 & {\bf 1}_3\\ -{\bf 1}_3 & 0 \end{array}\right)$. Near $x=0$, one can evaluate $\partial_xK, \tau$ and $\partial_xW_{c.s.}$ - this is done in appendix B.

Using (\ref{eq:Wconfl11}) - (\ref{eq:Wconfl12}) and (\ref{eq:Wconfl13})-(\ref{eq:Xisdefs}), one gets the following (near $x=0,\rho - \rho_0=0$):
\begin{equation}
\label{eq:Wconfl161}
D_xW_{c.s.}\approx ln x\Biggl({\cal A}_1 + {\cal B}_1x + {\cal C}_1 x ln x + {\cal D}_1 (\rho - \rho_0)
+ {\cal B}_1^\prime{\bar x} + {\cal C}_1^\prime{\bar x} ln {\bar x} + {\cal D}_1^\prime ({\bar\rho} - {\bar\rho_0})\Biggr)=0.
\end{equation}
Similarly,
\begin{equation}
\label{eq:Wconfl17}
D_{\rho - \rho_0}W_{c.s.}\approx {\cal A}_2 + {\cal B}_2 x + {\cal C}_2 x ln x + {\cal D}_2 (\rho - \rho_0)
+ {\cal B}_2^\prime{\bar x} + {\cal C}_2^\prime{\bar x} ln {\bar x} + {\cal D}_2^\prime ({\bar\rho} - {\bar\rho_0})=0.
\end{equation}

\item
$\underline{{\rm Near}\ \rho^6 + \phi= 1}$

Near $y\equiv \rho^6 + \rho - 1 =0$ and a small $\rho=\rho_0^\prime$, one can follow a similar analysis
as (\ref{eq:Wconfl11} - (\ref{eq:Wconfl17}) and arrive at similar equations:
\begin{eqnarray}
\label{eq:Wconfl21}
& & D_yW_{c.s.}\approx ln y\Biggl({\cal A}_3 + {\cal B}_3y + {\cal C}_3 y ln y + {\cal D}_3 (\rho - \rho_0^\prime)
+ {\cal B}_3^\prime{\bar y} + {\cal C}_3^\prime{\bar y} ln {\bar y} + {\cal D}_3^\prime ({\bar\rho} - {\bar\rho_0^\prime})\Biggr)=0,\nonumber\\
& & D_{\rho - \rho_0^\prime}W_{c.s.}\approx {\cal A}_4 + {\cal B}_4 y + {\cal C}_4 y ln y + {\cal D}_4 (\rho - \rho_0^\prime)
+ {\cal B}_4^\prime{\bar y} + {\cal C}_4^\prime{\bar y} ln {\bar y} + {\cal D}_4^\prime ({\bar\rho} - {\bar\rho_0^\prime})=0.
\end{eqnarray}

\item
$\underline{\rm Points\ away\ from\ both\ conifold\ loci}$

It can be shown, again following an analysis similar to the one carried out in (\ref{eq:Wconfl11}) -
(\ref{eq:Wconfl21}), that one gets the following set of equations from extremization of the complex-structure moduli superpotential:
\begin{equation}
\label{eq:Wawayconfl}
{\cal A}_i + {\cal B}_i\psi + {\cal C}_i\phi + {\cal B^\prime}_i{\bar\psi} + {\cal C^\prime}_i{\bar\phi} = 0,
\end{equation}
where $i$ indexes the different regions in the moduli space away from the two conifold loci, as discussed in section {\bf 2} earlier.

\end{itemize}

Therefore, to summarize,
\begin{eqnarray}
\label{eq:DW=0}
& & \underline{{\rm Near}\ \phi=\omega^{-1}}:\nonumber\\
& & {\cal A}_1 + {\cal B}_1 (\phi_1 - \omega^{-1}) + {\cal C}_1 (\phi_1 - \omega^{-1}) ln (\phi_1 - \omega^{-1}) + {\cal D}_1 (\rho_1 - \rho_0)\nonumber\\
& & + {\cal B}_1^\prime({\bar\phi_1} - \omega) + {\cal C}_1^\prime({\bar\phi} - \omega) ln ({\bar\phi} - \omega) + {\cal D}_1^\prime ({\bar\rho_1} - {\bar\rho_0})=0,\nonumber\\
& & {\cal A}_2 + {\cal B}_2 (\phi_1 - \omega^{-1}) + {\cal C}_2 (\phi_1 - \omega^{-1})
ln (\phi_1 - \omega^{-1}) + {\cal D}_2 (\rho_1 - \rho_0)\nonumber\\
& & + {\cal B}_2^\prime({\bar\phi_1} - \omega) + {\cal C}_2^\prime({\bar\phi_1} - \omega) ln ({\bar\phi_1} - \omega) + {\cal D}_2^\prime ({\bar\rho_1} - {\bar\rho_0})=0,\nonumber\\
& & \tau_1=\frac{\Xi[f_i;{\bar\phi_1} - \omega,\rho_1 - \rho_0]}{\sum_{i=0}^5h_i{\bar A_i}}\left[1 - \frac{\Xi[h_i;{\bar\phi_1} - \omega,\rho_1 - \rho_0]}{\sum_{j=0}^5h_i{\bar A_i}}\right];\nonumber\\
& & \underline{{\rm Near}\ \rho^6 + \phi - 1 = 0}:\nonumber\\
& & {\cal A}_3 + {\cal B}_3 (\rho_2^6+\phi-1) + {\cal C}_3 (\rho_2^6+\phi_2-1) ln (\rho_2^6+\phi_2-1) + {\cal D}_3 \phi_2
+ {\cal B}_3^\prime({\bar\rho_2^6}+{\bar\phi}-1)\nonumber\\
 & & + {\cal C}_3^\prime({\bar\rho_2^6}+{\bar\phi}-1) ln ({\bar\rho_2^6}+{\bar\phi}-1) + {\cal D}_3^\prime{\bar\phi}=0,\nonumber\\
& &  {\cal A}_4 + {\cal B}_4 (\rho_2^6+\phi_2-1) + {\cal C}_4 (\rho_2^6+\phi_2-1)
ln (\rho_2^6+\phi_2-1) + {\cal D}_4 \phi_2\nonumber\\
& & + {\cal B}_4^\prime({\bar\rho}^6-{\bar\phi}-1) + {\cal C}_4^\prime({\bar\rho}^6+{\bar\phi}-1) ln ({\bar\rho}^6+{\bar\phi}-1)+ {\cal D}_4^\prime {\bar\phi_2}=0,\nonumber\\
& & \tau_2=\frac{\Xi[f_i;{\bar\rho}^6+{\bar\phi}-1,\phi_2 ]}{\sum_{i=0}^5h_i{\bar A_i^\prime}}\left[1 - \frac{\Xi[h_i;{\bar\rho}^6+{\bar\phi}-1,\phi_2 ]}{\sum_{j=0}^5h_i{\bar A_i^\prime}}\right]\nonumber\\
& & \underline{|\phi^3|<1,\ {\rm Large}\ \psi}:\nonumber\\
& & {\cal A}_5 + {\cal B}_5(\phi_3 - \phi_0^{\prime\prime}) + {\cal C}_5(\rho_3 - \rho_0^{\prime\prime})
{\cal B}^\prime_5({\bar\phi}_3 - {\bar\phi}_0^{\prime\prime})
+ {\cal C}^\prime_5({\bar\rho}_3 - {\bar\rho}_0^{\prime\prime})= 0,\nonumber\\
& & {\cal A}_6 + {\cal B}_6(\phi_3 - \phi_0^{\prime\prime}) + {\cal C}_6(\rho_3 - \rho_0^{\prime\prime})
+ {\cal B}^\prime_6({\bar\phi}_3 - {\bar\phi}_0^{\prime\prime})
+ {\cal C}^\prime_6({\bar\rho}_3 - {\bar\rho}_0^{\prime\prime}) = 0,\nonumber\\
& & \tau_3=\frac{\tilde{\Xi}[f_i;{\bar\phi_3},\rho_3 ]}{\sum_{i=0}^5h_i{\bar A_i^{\prime\prime}}}\left[1 - \frac{\tilde{\Xi}[h_i;{\bar\phi_3},\rho_3]}{\sum_{j=0}^5h_i{\bar A_i^{\prime\prime}}}\right];\nonumber\\
& & \underline{\left|\frac{\rho^6}{\phi - \omega^{0,-1,-2}}\right|<1}:\nonumber\\
& & {\cal A}_7 + {\cal B}_7(\phi_4 - \phi_0^{\prime\prime\prime}) + {\cal C}_7(\rho_4 - \rho_0^{\prime\prime\prime})
+ {\cal B}^\prime_7({\bar\phi}_4 - {\bar\phi}_0^{\prime\prime\prime}) + {\cal C}^\prime_7({\bar\rho}_4 - {\bar\rho}_0^{\prime\prime\prime}) = 0,\nonumber\\
& & {\cal A}_7 + {\cal B}_7(\phi_4 - \phi_0^{\prime\prime\prime}) + {\cal C}_7(\rho_4 - \rho_0^{\prime\prime\prime})
+ + {\cal B}^\prime_7({\bar\phi}_4 - {\bar\phi}_0^{\prime\prime\prime})
+ {\cal C}^\prime_7({\bar\rho}_4 - {\bar\rho}_0^{\prime\prime\prime}) = 0,\nonumber\\
& & \tau_4=\frac{\tilde{\Xi}[f_i;{\bar\phi_4},\rho_4 ]}{\sum_{i=0}^5h_i{\bar A_i^{\prime\prime\prime}}}\left[1 - \frac{\tilde{\Xi}[h_i;{\bar\phi_4},\rho_4]}{\sum_{j=0}^5h_i{\bar A_i^{\prime\prime\prime}}}\right],\nonumber\\
\end{eqnarray}
where on deleting the $ln$ terms in $\Xi$ one gets the form of $\tilde{\Xi}$ in (\ref{eq:DW=0}).
 Given that the Euler characteristic of
the elliptically-fibered Calabi-Yau four-fold to which, according to the Sen's construction
\cite{SenFIIBorient}, the orientifold of the Calabi-Yau three-fold of (\ref{eq:hypersurface}) corresponds to,
will be very large\footnote{See \cite{Denef+Douglas+Florea} - $\chi(CY_4)=6552$ where the $CY_4$ for the ${\bf WCP^4}[1,1,1,6,9]$-model,
is the resolution of a
Weierstrass over a three-fold B with $D_4$ and $E_6$ singularities along two sections, with the three-fold
a ${\bf CP^1}$-fibration over ${\bf CP^2}$ with the two divisors contributing to the instanton superpotential a la Witten being
sections thereof.}, and further assuming the absence of
$D3$-branes, this would imply that one is allowed to take a large value of $f^T.\Sigma.h$, and hence
the fluxes - therefore, similar to the philosophy of \cite{VafaInverse}, we would disregard the integrality of fluxes. Without doing the numerics, we will now give a plausibility argument about the existence of solution to any one of the four sets of equations in (\ref{eq:DW=0}). As one can drop $x$ as compared to $x ln x$ for $x\sim0$, the equations in (\ref{eq:DW=0}) pair off either as:
\begin{itemize}
\item
Near either of the two conifold loci:
\begin{eqnarray}
\label{eq:DWsimp1}
& & A_i + (B_i cos \alpha_i + B_i^\prime sin \alpha_i) \epsilon_i ln \epsilon_i + C_i\beta_i + C_i^\prime {\bar\beta_i} = 0,\nonumber\\
& & \tilde{a}_i + (\tilde{b}_i cos \alpha_i + \tilde{b}_i^\prime sin \alpha_i) \epsilon_i ln \epsilon_i + \tilde{c}_i\beta_i + \tilde{c}_i^\prime {\bar\beta_i} = 0,
\end{eqnarray}
or
\item
Away from both the conifold loci:
\begin{eqnarray}
\label{eq:DWsimp2}
& & A_i + B_i \gamma_i + C_i\delta_i + B_i^\prime {\bar\gamma}_i + C_i^\prime {\bar\delta}_i = 0,\nonumber\\
& & \tilde{A}_i + \tilde{B}_i \gamma_i + \tilde{C}_i\delta_i + \tilde{B}_i^\prime {\bar\gamma}_i
+ \tilde{C}_i^\prime {\bar\delta}_i = 0,
\end{eqnarray}
where $\epsilon_i,\alpha_i$ correspond to the magnitude and phase of the extremum values of either $\phi - \omega^{-1}$ or $\rho^6 + \phi - 1$, and $\gamma_i,\delta_i$ are different (functions of) extremum values of $\phi,\psi$ near and away, respectively, from the two conifold loci, and both sets are understood to be ``close to zero" each.
 \end{itemize}
 From the point of view of practical calculations, let us rewrite, e.g., (\ref{eq:DWsimp1}) as the equivalent four real equations:
 \begin{eqnarray}
 \label{eq:DWsimp12}
 & & {\cal A}_i + {\cal B}_i cos\alpha_i \epsilon_i ln\epsilon_i + {\cal B}_i^\prime sin\alpha_i \epsilon_i ln\epsilon_i + {\cal C}_i Re(\beta_i) + {\cal C}_i^\prime Im(\beta_i) = 0,\nonumber\\
 & & \widetilde{{\cal A}_i} + \widetilde{{\cal B}_i} cos\alpha_i \epsilon_i ln\epsilon_i +
 \widetilde{{\cal B}_i}^\prime sin\alpha_i \epsilon_i ln\epsilon_i + \widetilde{{\cal C}_i} Re(\beta_i) + \widetilde{\cal C}_i^\prime Im(\beta_i) = 0,\nonumber\\
 & & {\nu}_i + {\chi}_i cos\alpha_i \epsilon_i ln\epsilon_i + {\chi}_i^\prime sin\alpha_i \epsilon_i ln\epsilon_i + {\vartheta}_i Re(\beta_i) + {\vartheta}^\prime Im(\beta_i) = 0,\nonumber\\
 & & \widetilde{{\nu}_i} + \widetilde{{\chi}_i} cos\alpha_i \epsilon_i ln\epsilon_i +
 \widetilde{{\chi}_i}^\prime sin\alpha_i \epsilon_i ln\epsilon_i + \widetilde{{\vartheta}_i} Re(\beta_i) + \widetilde{\vartheta}_i^\prime Im(\beta_i) = 0.
 \end{eqnarray}
  In (\ref{eq:DW=0}), by ``close to zero", what we would be admitting are, e.g., $\epsilon_i,|\beta_i|\sim e^{-5}\approx 7\times10^{-3}$ implying that
 $\epsilon_i ln\epsilon_i\approx10^{-2}$. Let us choose the moduli-independent constants in (\ref{eq:DWsimp12}),
  after suitable rationalization, to be $7\times {\cal O}(1)$, the coefficients of the
 $\epsilon_i ln\epsilon_i$-terms to be $7\times10^2$ and the coefficients of $Re(\beta_i)$ and $Im(\beta_i)$
 to be $\sim10^3$. On similar lines, for (\ref{eq:DWsimp2}), we could
  take the moduli to be $\sim e^{-5}$ and the moduli-independent and moduli-dependent constants to be
  $7\times {\cal O}(1)$ and $\sim10^3$ respectively. Now, the constants appearing in (\ref{eq:DWsimp12})
  (and therefore (\ref{eq:DW=0})) are cubic in the fluxes (more precisely, they are of the type
  $h^2f$ in obvious notations), which for  (\ref{eq:hypersurface}) would be $\sim10^3$ (See \cite{Denef+Douglas+Florea}).
  In other words, {\it for the same choice of the NS-NS and RR fluxes} - 12 in number - one gets 6 or 9 or 12 complex
(inhomogenous [in $\psi,\phi$] algebraic/transcendetal) constraints (coming from (\ref{eq:DW=0}))
on the 6 or 9 or 12 extremum values of the complex structure moduli ($\phi_i,\psi_i,\tau_i;\ i=1,2,3,4$) finitely separated from each other in the moduli space. In principle, as long as one keeps $f^T.\Sigma.h$ fixed, one should be able to tune the fluxes $f_i,h_i;\ i=0,...,5$ to be able to solve these equations.
  Therefore, the expected estimates of the values of the constants and the moduli tuned by the
  algebraic-geometric inputs of the periods in the different regions of the moduli space as
  discussed in section {\bf 2}, are reasonable implying the possibility of existence of ``area codes",
  and the interpolating domain walls \cite{CDGKL}. Of course, complete numerical calculations,
  which will be quite involved, will be needed to see explicitly everything working out.

\section{Non-supersymmetric dS minimum via Non-perturbative $\alpha^\prime$- and Instanton Corrections}

In this section, using the results of \cite{Grimm}, we show that after inclusion of non-perturbative
$\alpha^\prime$-corrections to the K\"{a}hler potential, in addition to the perturbative $\alpha^\prime$ corrections of \cite{BBHL},
as well as the non-perturbative instanton contributions to the superpotential, it may be possible to obtain a large volume
non-supersymmetric dS minimum (analogous to \cite{Balaetal2} for the non-supersymmetric AdS minimum) {\it without the addition of $\overline{D3}$-branes} - see
also \cite{westphal}.

Let us begin with a summary of the inclusion of perturbative $\alpha^\prime$-corrections to the K\"{a}hler potential in type IIB string theory
compactified on Calabi-Yau three-folds with NS-NS and RR fluxes turned on, as discussed in \cite{BBHL}. The
$(\alpha^\prime)^3-$ corrections contributing to the K\"{a}hler moduli space metric are contained in
\begin{equation}
\label{eq:nonpert1}
\int d^{10}x\sqrt{g}e^{-2\phi}\left(R + (\partial\phi)^2 + (\alpha^\prime)^3\frac{\zeta(3) J_0}{3.2^{11}}
+ (\alpha^\prime)^3(\bigtriangledown^2\phi) Q\right),
\end{equation}
where
\begin{eqnarray}
\label{eq:nonpert2}
& &  J_0\equiv t^{M_1N_1...M_4N_4}t_{M_1^\prime N_1^\prime....M_4^\prime N_4^\prime}
R^{M_1N_1}_{\ \ \ \ \ M_1^\prime N_1^\prime}...R^{M_4N_4}_{\ \ \ \ \ M_4^\prime N_4^\prime}\nonumber\\
& & + \frac{1}{4}\epsilon^{ABM_1N_1...M_4N_4}\epsilon_{ABM_1^\prime N_1^\prime...M_4^\prime N_4^\prime}
R^{M_1^\prime N_1^\prime}_{\ \ \ \ \ M_1 N_1}...R^{M_4^\prime N_4^\prime}_{\ \ \ \ \ M_4 N_4},
\end{eqnarray}
the second term in (\ref{eq:nonpert2}) being the ten-dimensional generalization of the eight-dimensional
Euler density, and
\begin{eqnarray}
\label{eq:nonpert3}
& &  t^{IJKLMNPQ}\equiv-\frac{1}{2}\epsilon^{IJKLMNPQ} -\frac{1}{2}\biggl[(\delta^{IK}\delta^{JL}
-\delta^{IL}\delta^{JK})(\delta^{MP}\delta^{NQ}-\delta^{MQ}\delta^{NP})\nonumber\\
& & +(\delta^{KM}\delta^{LN}-\delta^{KN}\delta^{LM})(\delta^{PI}\delta^{QJ}-\delta^{PJ}\delta^{QI})
+(\delta^{IM}\delta^{JN}-\delta^{IN}\delta^{JM})(\delta^{KP}\delta^{LQ}-\delta^{KQ}\delta^{LP})
\biggr]\nonumber\\
& & +\frac{1}{2}\left[\delta^{JK}\delta^{KM}\delta^{NP}\delta^{QI}+\delta^{JM}\delta^{NK}\delta^{LP}\delta^{QI}
+\delta^{JM}\delta^{NP}\delta^{QK}\delta^{KI}\right]\nonumber\\
& & {\rm +\ 45\ terms\ obtained\ by\ antisymmetrization\ w.r.t.}\ (ij),(kl),(mn),(pq),
\end{eqnarray}
and
\begin{equation}
\label{eq:nonpert31}
Q\equiv\frac{1}{12(2\pi)^3}\biggl(R_{IJ}^{\ \ }R_{KL}^{\ \ MN}R_{MN}^{\ \ IJ}
-2 R_{I\ K}^{\ K\ L}R_{K\ L}^{\ M\ N}R_{M\ N}^{\ I\ J}\biggr).
\end{equation}
The perturbative world-sheet corrections to the hypermultiplet moduli space of Calabi-Yau three-fold compactifications of
type II theories are captured by the prepotential:
\begin{equation}
\label{eq:nonpert4}
F(X)=\frac{i}{3}\kappa_{abc}\frac{X^aX^bX^c}{X^0} + (X^0)^2
\xi,
\end{equation}
where the $(\alpha^\prime)^3$-corrections are contained in
$\xi\equiv-(\alpha^\prime)^3\frac{\chi(CY_3)\zeta(3)}{2}$, $\kappa_{abc}$ being the classical $CY_3$ intersection numbers.
Substituting (\ref{eq:nonpert4}) in $K=-ln\left[X^i{\bar F}_i + {\bar X}^iF_i\right]$ gives:
\begin{equation}
\label{eq:nonpert5}
K = - ln\left[-\frac{i}{6}(z^a - {\bar z}^a)(z^b - {\bar z}^b)(z^c - {\bar z}^c) + 4\xi\right].
\end{equation}
Truncation of ${\cal N}=2$ to ${\cal N}=1$, implying reduction of the quaternionic geometry to K\"{a}hler
geometry, corresponds to a K\"{a}hler metric which becomes manifest in K\"{a}hler coordinates: $T^a=\frac{1}{3}g^a+i\hat{V}^a,
\tau=l+ie^{-\phi_0}$, the hat denoting the Einstein's frame in which, e.g.,
$\hat{V}_a=e^{\phi_0}\left(\frac{1}{6}\kappa_{abc}v^bv^c\right)$, $v^a$ being the K\"{a}hler moduli, and
the K\"{a}hler potential is given by:
\begin{equation}
\label{eq:nonpert6}
K = - ln\left(-(\tau-{\bar\tau})\right) - 2 ln\left(\hat{{\cal V}} + \frac{1}{2}\xi e^{{-3\phi_0}{2}}\right) -
ln\left(-i\int_{CY_3}\Omega\wedge{\bar\Omega}\right),
\end{equation}
substituting which into the ${\cal N}=1$ potential
$V = e^K\left(g^{i{\bar j}}D_iW{\bar D}_{\bar j}{\bar W} - 3 |W|^2\right)$ (one sums over all the moduli),
one gets:
\begin{eqnarray}
\label{eq:nonpert7}
& & V = e^K\Biggl[(G^{-1})^{\alpha{\bar\beta}}D_\alpha W D_{\bar\beta}{\bar W} + (G^{-1})^{\tau{\bar\tau}}D_\tau W
D_{\bar\tau}{\bar W}
- \frac{9\hat{\xi}\hat{{\cal V}}e^{-\phi_0}}{(\hat{\xi}-\hat{{\cal V}})(\hat{\xi} + 2\hat{{\cal V}})}(W{\bar D}_{\bar\tau}{\bar W}
+ {\bar W}D_\tau W)\nonumber\\
& & -3\hat{\xi}\frac{((\hat{\xi}) ^2 + 7\hat{\xi}\hat{{\cal V}} + (\hat{{\cal V}})^2)}{(\hat{\xi}-\hat{{\cal V}})
(\hat{\xi} + 2\hat{{\cal V}})^2}|W|^2\Biggr],
\end{eqnarray}
the hats being indicative of the Einstein frame - in our subsequent discussion, we will drop the hats for
notational convenience.
The structure of the $\alpha^\prime$-corrected potential shows that the no-scale structure is no longer preserved
due to explicit dependence of $V$ on $\hat{{\cal V}}$ and the $|W|^2$ term is not cancelled. In what follows,
we will be setting $2\pi\alpha^\prime=1$.

The type IIB Calabi-Yau orientifolds containing O3/O7-planes considered involve modding out by $(-)^{F_L}\Omega\sigma$ where
${\cal N}=1$ supersymmetry requires $\sigma$ to be a holomorphic and isometric involution:
$\sigma^*(J)=J,\ \sigma^*(\Omega)=-\Omega$. Writing the complexified K\"{a}hler form
$-B_2+iJ=t^A\omega=-b^a\omega_a+iv^\alpha\omega_\alpha$ where $(\omega_a,\omega_\alpha)$ form canonical
bases for ($H^2_-(CY_3,{\bf Z}), H^2_+(CY_3,{\bf Z})$), the $\pm$ subscript
indicative of being odd under $\sigma$, one sees that in the large volume limit of $CY_3/\sigma$,
contributions from large $t^\alpha=v^\alpha$ are exponentially suppressed, however the contributions
from $t^a=-B_a$ are not. Note that it is understood that $a$ indexes the {\bf real} subspace of {\bf real} dimensionality $h^{1,1}_-=2$; the  {\bf complexified} K\"{a}hler moduli correspond to $H^{1,1}(CY_3)$ with {\bf complex} dimensionality $h^{1,1}=2$ or equivalently real dimensionality equal to 4. So, even though $G^a=c^a-\tau b^a$ (for real $c^a$ and $b^a$ and complex $\tau$) is complex, the number of $G^a$'s is indexed by $a$ which runs over the real subspace $h^{1,1}_-(CY_3)$\footnote{To make the idea more explicit, the involution $\sigma$ under which the NS-NS two-form $B_2$ and the RR two-form $C_2$ are odd can be implemented as follows. Let $z_i, {\bar z}_i, i=1,2,3$ be the complex coordinates and the action of $\sigma$ be defined as: $z_1\leftrightarrow z_2, z_3\rightarrow z_3$; in terms of the $x_i$ figuring in the defining hypersurface in equation (1) on page 2, one could take $z_{1,2}=\frac{x_{1,2}^9}{x_5}$, etc. in the $x_5\neq0$ coordinate patch. One can construct the following bases $\omega^{(\pm)}$ of real two-forms of $H^2$ even/odd under the involution $\sigma$:
\begin{eqnarray}
\label{eq:bases}
& & \omega^{(-)}=\{\sum(dz^1\wedge d{\bar z}^{\bar 2} - dz^2\wedge d{\bar z}^{\bar 1}),
 i(dz^1\wedge d{\bar z}^{\bar 1} - dz^2\wedge d{\bar z}^{\bar 2})\}\equiv\{\omega^{(-)}_1,\omega^{(-)}_2\},\nonumber\\
& & \omega^{(+)}=\{\sum i(dz^1\wedge d{\bar z}^{\bar 2} + dz^2\wedge d{\bar z}^{\bar 1}),\sum i dz^1\wedge d{\bar z}^{\bar 1}\}\equiv\{\omega^{(+)}_1,\omega^{(+)}_2\}.
\end{eqnarray}
This implies that $h^{1,1}_+(CY_3)=h^{1,1}_-(CY_3)=2$ - the two add up to give 4 which is the {\bf real} dimensionality of $H^2(CY_3)$ for the given Swiss Cheese Calabi-Yau. As an example, let us write down $B_2\in{\bf R}$ as
\begin{eqnarray}
\label{eq:Bform}
B_2 & = & B_{1{\bar 2}}dz^1\wedge d{\bar z}^{\bar 2} + B_{2{\bar 3}}dz^2\wedge d{\bar z}^{\bar 3} + B_{3{\bar 1}}dz^3\wedge d{\bar z}^{\bar 1}  + B_{2{\bar 1}}dz^2\wedge d{\bar z}^{\bar 1} + B_{3{\bar 2}}dz^3\wedge d{\bar z}^{\bar 2} + B_{1{\bar 3}}dz^1\wedge d{\bar z}^{\bar 3}\nonumber\\
& & + B_{1{\bar 1}}dz^1\wedge d{\bar z}^{\bar 1}+ B_{2{\bar 2}}dz^2\wedge d{\bar z}^{\bar 2}+ B_{3{\bar 3}}dz^3\wedge d{\bar z}^{\bar 3}.
\end{eqnarray}
Now, using
(\ref{eq:bases}), one sees that by assuming $B_{1{\bar 2}}=B_{2{\bar 3}}=B_{3{\bar 1}}=b^1$, and $B_{1{\bar 1}}=-B_{2{\bar 2}}= i b^2, B_{3{\bar 3}}=0$, one can write $B_2=b^1\omega^{(-)}_1 + b^2\omega^{(-)}_2\equiv\sum_{a=1}^{h^{1,1}_-=2}b^a\omega^{(-)}_a$.}
As shown in \cite{Grimm}, based on the $R^4$-correction to the $D=10$ type IIB supergravity action
\cite{Green+Gutperle}
and the modular completion of ${\cal N}=2$ quaternionic geometry by summation over all $SL(2,{\bf Z})$ images
of world sheet corrections as discussed in \cite{Llanaetal},
the non-perturbative large-volume  $\alpha^\prime$-corrections that
survive the process of orientifolding of type IIB theories (to yield ${\cal N}=1$) to the
K\"{a}hler potential is given by (in the Einstein's frame):
\begin{eqnarray}
\label{eq:nonpert8}
& & K = - ln\left(-i(\tau-{\bar\tau})\right) - 2 ln\Biggl[{\cal V} + \frac{\chi}{2}\sum_{m,n\in{\bf Z}^2/(0,0)}
\frac{({\bar\tau}-\tau)^{\frac{3}{2}}}{(2i)^{\frac{3}{2}}|m+n\tau|^3}\nonumber\\
& &  - 4\sum_{\beta\in H_2^-(CY_3,{\bf Z})} n^0_\beta\sum_{m,n\in{\bf Z}^2/(0,0)}
\frac{({\bar\tau}-\tau)^{\frac{3}{2}}}{(2i)^{\frac{3}{2}}|m+n\tau|^3}cos\left((n+m\tau)k_a\frac{(G^a-{\bar G}^a)}{\tau - {\bar\tau}}
 - mk_aG^a\right)\Biggr],
\end{eqnarray}
where $n^0_\beta$ are the genus-0 Gopakumar-Vafa invariants for the curve $\beta$ and
$k_a=\int_\beta\omega_a$,  ,
and $G^a=c^a-\tau b^a$, the real RR two-form potential $C_2=C_a\omega^a$ and the real NS-NS two-form potential
$B_2=B_a\omega^a$. As pointed out in \cite{Grimm}, in (\ref{eq:nonpert8}),
one should probably sum over the orbits of the discrete
subgroup to which the symmetry group $SL(2,{\bf Z})$ reduces. Its more natural to write out the K\"{a}hler potential
and the superpotential in terms of the ${\cal N}=1$ coordinates $\tau, G^a$ and $T_\alpha$ where
\begin{equation}
\label{eq:nonpert10}
T_\alpha = \frac{i}{2}e^{-\phi_0}\kappa_{\alpha\beta\gamma}v^\beta v^\gamma - (\tilde{\rho}_\alpha - \frac{1}{2}\kappa_{\alpha ab}c^a b^b)
-\frac{1}{2(\tau - {\bar\tau})}\kappa_{\alpha ab}G^a(G^b - {\bar G}^b),
\end{equation}
where $\tilde{\rho}_\alpha$ being defined via $C_4$(the RR four-form potential)$=\tilde{\rho}_\alpha\tilde{\omega}_\alpha,
\tilde{\omega}_\alpha\in H^4_+(CY_3,{\bf Z})$.

Based on the action for the Euclidean $D3$-brane world volume (denoted by $\Sigma_4$) action \\
$iT_{D3}\int_{\Sigma_4} e^{-\phi}\sqrt{g-B_2+F}+T_{D3}\int_{\Sigma_4}e^C\wedge e^{-B_2+F}$, the nonperturbative
superpotential coming from a $D3$-brane wrapping a divisor $\Sigma\in H^4(CY_3/\sigma,{\bf Z})$ such that the unit
arithmetic genus condition of Witten \cite{Witten} is satisfied, will be proportional to
(See \cite{Grimm})
\begin{equation}
\label{eq:nonpertW1}
e^{\frac{1}{2}\int_\Sigma e^{-\phi}(-B_2+iJ)^2-i\int_\Sigma(C_4-C_2\wedge B_2+\frac{1}{2}C_0B_2^2}
= e^{iT_\alpha\int_\Sigma\tilde{\omega}_\alpha}\equiv e^{in_\Sigma^\alpha T_\alpha},
\end{equation}
where $C_{0,2,4}$ are the RR potentials. The prefactor multiplying (\ref{eq:nonpertW1}) is assumeto factorize
into a function of the ${\cal N}=1$ coordinates $\tau,G^a$ and a function of the other moduli. Based on
appropriate transformation properties of the superpotential under the shift symmetry and
$\Gamma_S\subset SL(2,{\bf Z})$:
\begin{eqnarray}
\label{eq:nonpert12}
& & (i)\ \tau\rightarrow\frac{a\tau + b}{c\tau + d},\nonumber\\
& & \left(\begin{array}{c} C_2\\B_2\end{array}\right)\rightarrow
\left(\begin{array}{cc} a & b \\ c & d \end{array}\right)
\left(\begin{array}{c} C_2\\B_2\end{array}\right),
\nonumber\\
& & G^a\rightarrow\frac{G^a}{(c\tau + d)},\nonumber\\
& & T^\alpha\rightarrow T_\alpha + \frac{c}{2}\frac{\kappa_{\alpha ab}G^a G^b}{(c\tau + d)};\nonumber\\
& & (ii)\ b^a\rightarrow b^a + 2\pi n^a,\nonumber\\
& & G^a\rightarrow G^a - 2\pi\tau n^a,\nonumber\\
& & T_\alpha\rightarrow T_\alpha - 2\pi\kappa_{\alpha ab}n^aG^b + 2\pi^2\tau\kappa_{\alpha ab}n^an^b,
\end{eqnarray}
the non-perturbative instanton-corrected superpotential was shown in \cite{Grimm} to be:
\begin{equation}
\label{eq:nonpert9}
W = \int_{CY_3}G_3\wedge\Omega + \sum_{n^\alpha}\frac{\theta_{n^\alpha}(\tau,G)}{f(\eta(\tau))}e^{in^\alpha T_\alpha},
\end{equation}
where the theta function is given as:
\begin{equation}
\label{eq:nonpert11}
\theta_{n^\alpha}(\tau,G)=\sum_{m_a}e^{\frac{i\tau m^2}{2}}e^{in^\alpha G^am_a}.
\end{equation}
In (\ref{eq:nonpert11}), $m^2=C^{ab}m_am_b, C_{ab}=-\kappa_{\alpha^\prime ab}$, $\alpha=\alpha^\prime$
corresponding to that $T_\alpha=T_{\alpha^\prime}$ (for simplicity) that is invariant under (\ref{eq:nonpert12}).

Now, for (\ref{eq:hypersurface}), as shown in \cite{Denef+Douglas+Florea}, there are two divisors which
when uplifted to an elliptically-fibered Calabi-Yau, have a unit arithmetic genus (\cite{Witten}):
$\tau_1\equiv\partial_{t_1}{\cal V}=\frac{t^2_1}{2},\ \tau_2\equiv\partial_{t_2}{\cal V}=\frac{(t_1+6t_2)^2}{2}$. In (\ref{eq:nonpert10}), $\rho_1=\tilde{\rho}_1-i\tau_1$ and
$\rho_2=\tilde{\rho}_2-i\tau_2$.

To set the notations, the metric corresponding to the K\"{a}hler potential in (\ref{eq:nonpert8}), will
be given as:
\begin{equation}
\label{eq:nonpert13}
G_{A{\bar B}}=\left(\begin{array}{cccc}
\partial_{\rho_1}{\bar\partial}_{{\bar\rho_1}}K & \partial_{\rho_1}{\bar\partial}_{{\bar\rho_2}}K
& \partial_{\rho_1}{\bar\partial}_{{\bar G}^1}K
& \partial_{\rho_1}{\bar\partial}_{{\bar G}^2}K\\
\partial_{\rho_2}{\bar\partial}_{{\bar\rho_1}}K & \partial_{\rho_2}{\bar\partial}_{{\bar\rho_2}}K
& \partial_{\rho_2}{\bar\partial}_{{\bar G}^1}K
& \partial_{\rho_2}{\bar\partial}_{{\bar G}^2}K\\
\partial_{G^1}{\bar\partial}_{{\bar\rho}_1}K & \partial_{G^1}{\bar\partial}_{{\bar\rho}_2}K &
\partial_{G^1}{\bar\partial}_{{\bar G}^1}K & \partial_{G^1}{\bar\partial}_{{\bar G}^2}K \\
\partial_{G^2}{\bar\partial}_{{\bar\rho}_1}K & \partial_{G^2}{\bar\partial}_{{\bar\rho}_2}K &
\partial_{G^2}{\bar\partial}_{{\bar G}^2}K & \partial_{G^2}{\bar\partial}_{{\bar G}^2}K
\end{array}\right),
\end{equation}
where $A\equiv\rho^{1,2},G^{1,2}$. We have taken the involution to be such that $h^{1,1}_-=2$. From the K\"{a}hler potential given in (\ref{eq:nonpert8}), one can show
that the corresponding K\"{a}hler metric of (\ref{eq:nonpert13}) is given by:
\begin{eqnarray}
\label{eq:nonpert131}
& & G_{A{\bar B}} =\nonumber\\
& & \left(\begin{array}{cccc}
\frac{1}{4}\left(\frac{1}{6\sqrt{2}}\frac{1}{\sqrt{{\bar\rho}_1 - \rho_1}{\cal Y}}
+ \frac{1}{18}\frac{({\bar\rho}_1 - \rho_1)}{{\cal Y}^2}\right)
&\frac{1}{144}\left(\frac{\sqrt{({\bar\rho}_1 - \rho_1)({\bar\rho}_2 - \rho_2)}}{{\cal Y}^2}\right)
& \frac{-ie^{-\frac{3\phi_0}{2}}\sqrt{{\bar\rho}_1 - \rho_1}{\cal Z}(\tau)}{6\sqrt{2}{\cal Y}^2}
& \frac{-ie^{-\frac{3\phi_0}{2}}\sqrt{{\bar\rho}_1 - \rho_1}{\cal Z}(\tau)}{6\sqrt{2}{\cal Y}^2}  \\
\frac{1}{144}\left(\frac{\sqrt{({\bar\rho}_1 - \rho_1)({\bar\rho}_2 - \rho_2)}}{{\cal Y}^2}\right)
& \frac{1}{4}\left(\frac{1}{6\sqrt{2}}\frac{\sqrt{{\bar\rho}_2 - \rho_2}}{{\cal Y}}
+ \frac{1}{18}\frac{\sqrt{{\bar\rho}_2 - \rho_2}}{{\cal Y}^2}\right)
& \frac{-ie^{-\frac{3\phi_0}{2}}\sqrt{{\bar\rho}_2 - \rho_2}{\cal Z}(\tau)}{6\sqrt{2}{\cal Y}^2}
& \frac{-ie^{-\frac{3\phi_0}{2}}\sqrt{{\bar\rho}_2 - \rho_2}{\cal Z}(\tau)}{6\sqrt{2}{\cal Y}^2}  \\
\frac{ik_1e^{-\frac{3\phi_0}{2}}\sqrt{{\bar\rho}_1 - \rho_1}{\cal Z}({\bar\tau})}{6\sqrt{2}{\cal Y}^2}
& \frac{ik_1e^{-\frac{3\phi_0}{2}}\sqrt{{\bar\rho}_2 - \rho_2}{\cal Z}({\bar\tau})}{6\sqrt{2}{\cal Y}^2}
& k_1^2{\cal X}_1 & k_1k_2{\cal X}_1 \\
\frac{ik_2e^{-\frac{3\phi_0}{2}}\sqrt{{\bar\rho}_1 - \rho_1}{\cal Z}({\bar\tau})}{6\sqrt{2}{\cal Y}^2}
& \frac{ik_2e^{-\frac{3\phi_0}{2}}\sqrt{{\bar\rho}_2 - \rho_2}{\cal Z}({\bar\tau})}{6\sqrt{2}{\cal Y}^2}
& k_1k_2 {\cal X}_1 & k_2^2{\cal X}_1
\end{array}\right), \nonumber\\
& &
\end{eqnarray}
where
\begin{eqnarray}
\label{eq:nonpert14}
& & {\cal Z}(\tau)\equiv \sum_c\sum_{m,n}A_{n,m,n_{k^c}}(\tau) sin(nk.b + mk.c),\nonumber\\
&& {\cal Y}\equiv {\cal V}_E + \frac{\chi}{2}\sum_{m,n\in{\bf Z}^2/(0,0)}
\frac{(\tau - {\bar\tau})^{\frac{3}{2}}}{(2i)^{\frac{3}{2}}|m+n\tau|^3} \nonumber\\
& &
- 4\sum_{\beta\in H_2(CY_3,{\bf Z})}n^0_\beta\sum_{m,n\in{\bf Z}^2/(0,0)}
\frac{(\tau - {\bar\tau})^{\frac{3}{2}}}{(2i)^{\frac{3}{2}}|m+n\tau|^3}cos\left((n+m\tau)k_a\frac{(G^a-{\bar G}^a)}{\tau - {\bar\tau}}
 - mk_aG^a\right),\nonumber\\
& & {\cal X}_1\equiv\frac{\sum_c\sum_{m,n\in{\bf Z}^2/(0,0)}e^{-\frac{3\phi_0}{2}}|n+m\tau|^3|
A_{n,m,n_{k^c}}(\tau)|^2cos(nk.b + mk.c)}{{\cal Y}}
 \nonumber\\
& & + \frac{|\sum_c\sum_{m,n\in{\bf Z}^2/(0,0)}e^{-\frac{3\phi_0}{2}}|n+m\tau|^3A_{n,m,n_{k^c}}(\tau)sin(nk.b + mk.c)|^2}{{\cal Y}^2},
\nonumber\\
& & A_{n,m,n_{k^c}}(\tau)\equiv \frac{(n+m\tau)n_{k^c}}{|n+m\tau|^3}.
\end{eqnarray}

The inverse metric is given as:
\begin{equation}
\label{eq:nonpert15}
G^{-1}=\left(\begin{array}{cccc}
(G^{-1})^{\rho_1{\bar\rho_1}} & (G^{-1})^{\rho_1{\bar\rho_2}} & (G^{-1})^{\rho_1{\bar G^1}} & 0 \\
\overline{(G^{-1})^{\rho_1{\bar\rho_2}}} & (G^{-1})^{\rho_2{\bar\rho_2}} & (G^{-1})^{\rho_2{\bar G^1}} & 0 \\
\overline{(G^{-1})^{\rho_1{\bar G^1}}} & \overline{(G^{-1})^{\rho_2{\bar G^1}}} & \frac{1}{(k_1^2-k_2^2){\cal X}_1}
& \frac{k_2}{(k_1k_2^2-k_1^3){\cal X}_1} \\
0 & 0 & \frac{k_2}{(k_1k_2^2-k_1^3){\cal X}_1} & \frac{1}{(k_1^2-k_2^2){\cal X}_1}
\end{array}\right),
\end{equation}
where the non-zero elements are given in appendix C.

Now, analogous to \cite{Balaetal2}, we will work in the large volume limit: ${\cal V}\rightarrow\infty,
\tau_1\sim ln {\cal V},\ \tau_2\sim {\cal V}^{\frac{2}{3}}$. In this limit, the inverse metric (\ref{eq:nonpert15})
simplifies to (we will not be careful about the magnitudes of the numerical factors in the following):
\begin{equation}
\label{eq:nonpert16}
G^{-1}\sim\left(\begin{array}{cccc}
-{\cal V}\sqrt{ln {\cal V}} & {\cal V}^{\frac{2}{3}}ln {\cal V}
& \frac{-i{\cal X}ln {\cal V}}{{\cal X}_2}& 0 \\
{\cal V}^{\frac{2}{3}}ln {\cal V} & {\cal V}^{\frac{4}{3}} &
\frac{i{\cal X}{\cal V}^{\frac{2}{3}}}{k_1{\cal X}_2}& 0\\
 \frac{i{\cal X} ln {\cal V}}{{\cal X}_2}
 & \frac{-i{\cal X}{\cal V}^{\frac{2}{3}}}{k_1{\cal X}_2}
& \frac{1}{(k_1^2-k_2^2){\cal X}_1}& \frac{k_2}{(k_1k_2^2-k_1^3){\cal X}_1} \\
0 & 0 & \frac{k_2}{(k_1k_2^2-k_1^3){\cal X}_1} & \frac{1}{(k_1^2-k_2^2){\cal X}_1}
\end{array}\right),
\end{equation}
where
$${\cal X}_2\equiv \sum_c\sum_{m,n\in{\bf Z}^2/(0,0)}|n+m\tau|^3|
A_{n,m,n_{k^c}}(\tau)|^2cos(nk.b + mk.c).$$ Refer to \cite{Balaetal2} for discussion on the minus sign in the $\left(G^{-1}\right)^{\rho_1{\bar\rho}_1}$.
Having extremized the superpotential w.r.t. the complex structure moduli and the axion-dilaton
modulus, the ${\cal N}=1$ potential
will be given by:
\begin{eqnarray}
\label{eq:nonpert17}
& & V = e^K\Biggl[\sum_{A,B=\rho_\alpha,G^a}\Biggl\{(G^{-1})^{A{\bar B}}\partial_A W_{np}{\bar\partial}_{\bar B}{\bar W_{np}}
+ \left((G^{-1})^{A{\bar B}}(\partial_A K){\bar\partial}_{\bar B}{\bar W_{np}}W + c.c.\right)\Biggr\} \nonumber\\
& & + \left(\sum_{A,B=\rho_\alpha,G^a}(G^{-1})^{A{\bar B}}\partial_A K{\bar\partial}_{\bar B}K - 3\right)|W|^2 + \sum_{\alpha,{\bar\beta}\in{\rm c.s.}}(G^{-1})^{\alpha{\bar\beta}}\partial_{\alpha} K_{c.s.}{\bar\partial}_{\bar\beta}K_{c.s.}|W_{np}|^2
\Biggr],
\end{eqnarray}
where the total superpotential $W$ is the sum of the complex structure moduli Gukov-Vafa-Witten superpotential $W_{c.s.}$
and the non-perturbative superpotential $W_{np}$ arising because of instantons (obtained by wrapping of
$D3$-branes around the divisors with complexified volumes $\tau_1$ and $\tau_2$).

Now, using:
\begin{eqnarray}
\label{eq:nonpert18}
& & \partial_{\rho_\alpha}W=\frac{\theta_{n^\alpha}(\tau,G^a)}{f(\eta(\tau))}e^{in^\alpha T_\alpha}in^\alpha
\left(-\frac{e^{-\phi}}{2}\right),\nonumber\\
& & \hskip -2cm
\partial_{G^a}W=\sum_{n^\alpha}\frac{e^{i\frac{\tau m^2}{2}}}{f(\eta(\tau))}e^{im_aG^an^\alpha}e^{in^\alpha T_\alpha}
\left(im_an^\alpha - in^\alpha\frac{\kappa_{\alpha ab}}{2|\tau-{\bar\tau}|^2}\left[{\bar\tau}(G^b-{\bar G}^b)
+({\bar\tau}G^b-\tau{\bar G}^b) + \frac{(2G^b - {\bar G}^b)}{(\tau-{\bar\tau})}\right]\right),
\end{eqnarray}
in the large-volume limit, one forms tables 1, 2 and 3.

One therefore sees from table 1 that the dominant term in $(G^{-1})^{A{\bar B}}\partial_A W_{np}{\bar\partial}_{\bar B}{\bar W_{np}}$ is $(G^{-1})^{\rho_1{\bar\rho}_1}|\partial_{\rho_1}W_{np}|^2$,
given by:
\begin{equation}
\label{eq:nonpert181}
\frac{{\cal Y}\sqrt{ln {\cal V}}}{{\cal V}^{2n^1e^{-\phi}}}e^{-2\phi}(n^1)^2\left|\frac{\theta_{n^1}(\tau,G)}{f(\eta(\tau))}
\right|^2e^{-2n^1 Im(T_1)}.
\end{equation}

From table 2 we see that the dominant term in $(G^{-1})^{A{\bar B}}(\partial_A K){\bar\partial}_{\bar B}{\bar W_{np}}W$ is $(G^{-1})^{\rho_1{\bar\rho}_1}\partial_{\rho_1}K{\bar\partial}_{\bar\rho_1}{\bar W}_{np}W+c.c.$,
which gives:
\begin{equation}
\label{eq:nonpert19}
\frac{W_{c.s.} ln {\cal V}}{{\cal V}^{n^1e^{-\phi}}}\left(\frac{\theta_{n^1}({\bar\tau},{\bar G})}{f(\eta({\bar\tau}))}
\right)e^{-in^1(-\tilde{\rho_1}+\frac{1}{2}\kappa_{1ab}
\frac{{\bar\tau}G^a-\tau{\bar G}^a}{({\bar\tau}-\tau)}\frac{(G^b-{\bar G}^b)}{({\bar\tau}-\tau)} -
\frac{1}{2}\kappa_{1ab}\frac{G^a(G^b-{\bar G}^b)}{(\tau-{\bar\tau})})}+c.c.
\end{equation}

Note, from table 3, the dominant and the (sub) sub-dominant terms in $(G^{-1})^{A{\bar B}}\partial_A K{\bar\partial}_{\bar B}K|W|^2$, given respectively by $(G^{-1})^{\rho_2{\bar\rho}_2}|\partial_{\rho_2}K|^2|W|^2$
and $\left[(G^{-1})^{\rho_1{\bar\rho}_1}|\partial_{\rho_1}K|^2 + (G^{-1})^{\rho_1{\bar\rho}_2}\partial_{\rho_1}{\bar\partial}_{\bar\rho_2}K + c.c.\right]|W|^2$
are actually of the form:
$\left[\frac{{\cal V}}{{\cal V} + \xi}\approx 1-\frac{\xi}{{\cal V}}+{\cal O}\left(\frac{1}{{\cal V}^2}\right)\right]|W|^2$ and
$\left[\frac{(ln {\cal V})^{\frac{3}{2}}}{({\cal V}+\xi)} + \frac{(ln {\cal V})^{\frac{3}{2}}}{({\cal V}+\xi)^2}\right]|W|^2$
$\approx\biggl[\frac{(ln {\cal V})^{\frac{3}{2}}}{{\cal V}}-\xi\frac{(ln {\cal V})^{\frac{3}{2}}}{{\cal V}^2}
+ \frac{(ln {\cal V})^{\frac{3}{2}}}{{\cal V}^2} -\xi\frac{(ln {\cal V})^{\frac{3}{2}}}{{\cal V}^3}$
$+{\cal O}\left(\frac{1}{{\cal V}^3}\right)\biggr]|W|^2$
respectively and the $\xi$-independent terms together cancel the ``-3" in (\ref{eq:nonpert16}). This is
just a rederivation of the last term in (\ref{eq:nonpert7}). One notes that there are additional terms
of ${\cal O}\left(\frac{1}{{\cal V}}\right)$ that one gets from
$\Biggl[(G^{-1})^{G^1{\bar G^1}}|\partial_{G^1}K|^2 + (G^{-1})^{G^2{\bar G^2}}|\partial_{G^2}K|^2 +
(G^{-1})^{G^1{\bar G^2}}\partial_{G^1}K{\bar\partial}_{\bar G^2}K\Biggr]|W|^2$, which is given by:
\begin{equation}
\label{eq:nonpert20}
\frac{|W|^2}{{\cal V}}\left(\frac{3k_2^2+k_1^2}{k_1^2-k_2^2}\right)
\frac{\left|\sum_c\sum_{n,m\in{\bf Z}^2/(0,0)}e^{-\frac{3\phi}{2}}A_{n,m,n_{k^c}}(\tau) sin(nk.b+mk.c)\right|^2}
{\sum_{c^\prime}\sum_{m^\prime,n^\prime\in{\bf Z}^2/(0,0)} e^{-\frac{3\phi}{2}}|n+m\tau|^3
|A_{n^\prime,m^\prime,n_{k^{c^{\prime}}}}(\tau)|^2 cos(n^\prime k.b+m^\prime k.c)},
\end{equation}
which one sees {\it can be either positive or negative}.

To summarize, from (\ref{eq:nonpert18}) - (\ref{eq:nonpert20}), one gets the following potential:
\begin{eqnarray}
\label{eq:nonpert21}
& & V\sim\frac{{\cal Y}\sqrt{ln {\cal V}}}{{\cal V}^{2n^1+2}}e^{-2\phi}(n^1)^2\frac{\left(\sum_{m^a}e^{-\frac{m^2}{2g_s} + \frac{m_ab^a n^1}{g_s} + \frac{n^1\kappa_{1ab}b^ab^b}{2g_s}}\right)^2}{\left|f(\eta(\tau))\right|^2}
\nonumber\\
& & +\frac{ln {\cal V}}{{\cal V}^{n^1+2}}\left(\frac{\theta_{n^1}({\bar\tau},{\bar G})}{f(\eta({\bar\tau}))}
\right)e^{-in^1(-\tilde{\rho_1}+\frac{1}{2}\kappa_{1ab}
\frac{{\bar\tau}G^a-\tau{\bar G}^a}{({\bar\tau}-\tau)}\frac{(G^b-{\bar G}^b)}{({\bar\tau}-\tau)} -
\frac{1}{2}\kappa_{1ab}\frac{G^a(G^b-{\bar G}^b)}{(\tau-{\bar\tau})})}+c.c.\nonumber\\
& & +
\frac{|W|^2}{{\cal V}^3}\left(\frac{3k_2^2+k_1^2}{k_1^2-k_2^2}\right)
\frac{\left|\sum_c\sum_{n,m\in{\bf Z}^2/(0,0)}e^{-\frac{3\phi}{2}}A_{n,m,n_{k^c}}(\tau) sin(nk.b+mk.c)\right|^2}
{\sum_{c^\prime}\sum_{m^\prime,n^\prime\in{\bf Z}^2/(0,0)} e^{-\frac{3\phi}{2}}|n+m\tau|^3
|A_{n^\prime,m^\prime,n_{k^{c^{\prime}}}}(\tau)|^2 cos(n^\prime k.b+m^\prime k.c)}+\frac{\xi|W|^2}{{\cal V}^3}.
\nonumber\\
& &
\end{eqnarray}
On comparing (\ref{eq:nonpert21}) with the analysis of \cite{Balaetal2}, one sees that for generic values of
the moduli $\rho_\alpha, G^a, k^{1,2}$ and ${\cal O}(1)\ W_{c.s.}$, {\it and $n^1=1$}, analogous to \cite{Balaetal2}, the second term
dominates; the third term is a new term. However, as in KKLT scenarios (See \cite{KKLT}), $W_{c.s.}<<1$; we would henceforth assume that the fluxes and complex structure moduli have been so fine tuned/fixed that $W\sim W_{n.p.}$. Further, from studies related to study of axionic slow roll inflation in Swiss Cheese models \cite{misra-shukla-II}, it becomes necessary to take $n^1>2$.  We assume that the fundamental-domain-valued $b^a$'s satisfy: $\frac{|b^a|}{\pi}<<1$\footnote{If one puts in appropriate powers of the Planck mass $M_p$, $\frac{|b^a|}{\pi}<<1$ is equivalent to $|b^a|<<M_p$, i.e., NS-NS axions are super sub-Planckian.}. This implies that the first term in (\ref{eq:nonpert21}) - $|\partial_{\rho^1}W_{np}|^2$ - a positive definite term and denoted henceforth by $V_I$, is the most dominant. Hence, if a minimum exists, it will be positive. To evaluate the extremum of the potential, one sees that:
\begin{eqnarray}
\label{eq:extrV-c_b}
& & \partial_{c^a}V_I\nonumber\\
& & \hskip -0.3in\sim- 4\frac{\sqrt{ln {\cal V}}}{{\cal V}^{2n^1+2}}\sum_{\beta\in H_2^-(CY_3,{\bf Z})} n^0_\beta\sum_{m,n\in{\bf Z}^2/(0,0)}mk^a
\frac{({\bar\tau}-\tau)^{\frac{3}{2}}}{(2i)^{\frac{3}{2}}|m+n\tau|^3} sin(n k.b + mk.c)\frac{\left(\sum_{m^a}e^{-\frac{m^2}{2g_s} + \frac{m_ab^a n^1}{g_s} + \frac{n^1\kappa_{1ab}b^ab^b}{2g_s}}\right)^2}{\left|f(\eta(\tau))\right|^2}=0\nonumber\\
& & \Leftrightarrow nk.b + mk.c = N\pi;\nonumber\\
& & \hskip-0.23in\partial_{b^a}V_I|_{nk.b + mk.c = N\pi}\sim\frac{{\cal V}\sqrt{ln {\cal V}}}{{\cal V}^{2n^1+1}}\frac{e^{-\frac{m^2}{2g_s} + \frac{m_{a^\prime}b^{a^\prime} n^1}{g_s} + \frac{n^1\kappa_{1a^\prime b^\prime}b^{a^\prime}b^{b^\prime}}{2g_s}}\sum_{m^a}e^{-\frac{m^2}{2g_s} + \frac{m_ab^a n^1}{g_s} + \frac{n^1\kappa_{1ab}b^ab^b}{2g_s}}}{\left|f(\eta(\tau))\right|^2}\left(\frac{n^1m^a}{g_s} + \frac{n^1\kappa_{1ab}b^b}{g_s}\right)=0.\nonumber\\
& & \end{eqnarray}
Now, given the ${\cal O}(1)$ triple-intersection numbers and super sub-Planckian NS-NS axions, we see that potential $V_I$ gets automatically extremized for $D1$-instanton numbers $m^a>>1$. Note that if the NS-NS axions get stabilized as per $\frac{n^1m^a}{g_s} + \frac{n^1\kappa_{1ab}b^b}{g_s}=0$, satisfying $\partial_{b^a}V=0$, this would imply that the NS-NS axions get stabilized at a rational number, and in particular, a value which is not a rational multiple of $\pi$, the same being in conflict with the requirement $nk.b + mk.c = N \pi$.
It turns out that the locus $nk.b + mk.c = N\pi$ for $|b^a|<<\pi$ and $|c^a|<<\pi$ corresponds to a flat saddle point with the NS-NS axions providing a flat direction - See \cite{misra-shukla-II}.

Analogous to \cite{Balaetal2}, for all directions in the moduli space with ${\cal O}(1)$ $W_{c.s.}$ and away from $D_iW_{cs}=D_\tau W=0=\partial_{c^a}V=\partial_{b^a}V=0$, the ${\cal O}(\frac{1}{{\cal V}^2})$ contribution
of $\sum_{\alpha,{\bar\beta}\in{c.s.}}(G^{-1})^{\alpha{\bar\beta}}D_\alpha W_{cs}{\bar D}_{\bar\beta}{\bar W}_{cs}$  dominates over (\ref{eq:nonpert21}),
ensuring that that there must exist a minimum, and given the positive definiteness of the potential $V_I$, this will be a dS minimum. There has been no need to add any $\overline{D3}$-branes as in KKLT to generate a dS vacuum. Also, interestingly, one can show that the condition $nk.b + mk.c = N \pi$ gurantees that the slow roll parameters ``$\epsilon$" and ``$\eta$" are much smaller than one for slow roll inflation beginning from the saddle point
and proceeding along an NS-NS axionic flat direction towards the nearest dS minimum (See \cite{misra-shukla-II}).

\section{The ``Inverse Problem" for Extremal Black Holes}

We now switch gears and address two issues in this and the subsequent sections, related to supersymmetric and non-supersymmetric black hole attractors\footnote{See \cite{Mohaupt} for a nice review of special geometry relevant to sections {\bf 5} and {\bf 6}.}. In this section,
using the techniques discussed in \cite{VafaInverse}, we explicitly solve the ``inverse problem" for extremal black holes in type II compactifications on (the mirror of) (\ref{eq:hypersurface}) - given a point in the moduli space, to find the charges $(p^I,q_I)$ that would satisfy $\partial_iV_{BH}=0$,
 $V_{BH}$ being the black-hole potential. In the next section, we address the issue of existence of ``fake superpotentials" in the same context.

 We will now summarize the ``inverse problem" as discussed in \cite{VafaInverse}).
 Consider $D=4, N=2$ supergravity coupled to $n_V$ vector multiplets in the absence of higher derivative
 terms. The black-hole potential can be written as \cite{nonsusybh1}:
 \begin{equation}
 \label{eq:BHinv1}
 V_{BH} = -\frac{1}{2}(q_I - {\cal N}_{IK} p^K)\left((Im {\cal N})^{-1}\right)^{IJ}(q_J - {\bar{\cal N}}p^L),
 \end{equation}
 where the $(n_V + 1)\times(n_V + 1)$ symmetric complex matrix, ${\cal N}_{IJ}$, the vector multiplet moduli space metric, is defined as:
 \begin{equation}
 \label{eq:BHinv2}
 {\cal N}_{IJ} \equiv {\bar F}_{IJ} + \frac{2i Im(F_{IK}) X^K Im (F_{IL}) X^L}{Im(F_{MN}) X^M X^N},
 \end{equation}
 $X^I,F_J$ being the symplectic sections and $F_{IJ}\equiv\partial_IF_J=\partial_JF_I$. The black-hole potential (\ref{eq:BHinv1}) can be rewritten (See \cite{VafaInverse}) as:
 \begin{equation}
 \label{eq:BHinv3}
 \tilde{V}_{BH} = \frac{1}{2}{\cal P}^I Im({\cal N}_{IJ}){\bar{\cal P}}^J - \frac{i}{2}{\cal P}^I(q_I - {\cal N}_{IJ}p^J)
 + \frac{i}{2}{\bar{\cal P}}^I(q_I - {\bar{\cal N}}_{IJ}p^J).
 \end{equation}
 The variation of (\ref{eq:BHinv3}) w.r.t. ${\cal P}^I$ gives:
 \begin{equation}
 \label{eq:BHinv4}
 {\cal P}^I=-i\left((Im {\cal N})^{-1})^{IJ}\right)(q_J - {\cal N}_{IJ}p^J),
 \end{equation}
 which when substituted back into (\ref{eq:BHinv3}), gives (\ref{eq:BHinv1}). From (\ref{eq:BHinv4}), one
 gets:
\begin{eqnarray}
\label{eq:BHinv5}
& & p^I = Re({\cal P}^I),\nonumber\\
& & q_I = Re({\cal N}_{IJ}{\cal P}^J).
\end{eqnarray}
Extremizing $\tilde{V}_{BH}$ gives:
\begin{equation}
\label{eq:BHinv6}
{\cal P}^I{\bar{\cal P}}^J\partial_i Im({\cal N}_{IJ}) + i({\cal P}^I\partial_i {\cal N}_{IJ} - {\bar{\cal P}}^J\partial_i{\bar{\cal N}}_{IJ})p^J = 0,
\end{equation}
which using (\ref{eq:BHinv5}) yields:
\begin{equation}
\label{eq:BHinv7}
\partial_i Im({\cal P}^I{\cal N}_{IJ}{\cal P}^J) = 0.
\end{equation}
Similar to what was done in section {\bf 3}, one uses the semi-classical approximation and
disregards the integrality of the electric and magnetic charges taking them to be large.

The inverse problem is not straight forward to define as all sets of charges $(p^I,q_I)$ which are related
to each other by an $Sp(2n_V + 2,{\bf Z})$-transformation, correspond to the same point in the moduli space. This is because the $V_{BH})$ (and $\partial_iV_{BH}$) is (are) symplectic invariants. Further, $\partial_iV_{BH}=0$ give $2n_V$ real equations in $2n_V+2$ real variables $(p^I,q_I)$. To fix these two
problems, one looks at critical values of $V_{BH}$ in a fixed gauge $W=w\in{\bf C}$. In other words,
\begin{equation}
\label{eq:BHinv8}
W=\int_M\Omega\wedge H = q_I X^I - p^I F_I = X^I(q_I - {\cal N}_{IJ}p^J) = w,
\end{equation}
which using (\ref{eq:BHinv5}), gives:
\begin{equation}
\label{eq:BHinv9}
X^I Im({\cal N}_{IJ}){\bar{\cal P}}^J = w.
\end{equation}
Thus, the inverse problem boils down to solving:
\begin{eqnarray}
\label{eq:BHinv10}
& & p^I = Re({\cal P}^I),\ q_I=Re({\cal N}_{IJ}{\cal P}^J);\nonumber\\
& & \partial_i({\cal P}^I{\cal N}_{IJ}{\cal P}^J)=0,\ X^I{\cal N}_{IJ}{\bar{\cal P}}^J = iw.
\end{eqnarray}
One solves for ${\cal P}^I$s from the last two equations in (\ref{eq:BHinv10}) and substitutes the result
into the first two equations of (\ref{eq:BHinv10}).

We will now solve the last two equations of (\ref{eq:BHinv10}) for (\ref{eq:hypersurface}). As an example,
we work with points in the moduli space close to one of the two conifold loci: $\phi^3=1$. We need to work out the matrix $F_{IJ}$ so that one can work out the matrix ${\cal N}_{IJ}$. From the symmetry of $F_{IJ}$
w.r.t. $I$ and $J$, one sees that the constants appearing in (\ref{eq:confl210}) must satisfy some
constraints (which must be borne out by actual numerical computations). To summarize, near $x=0$ and using
(\ref{eq:confl24})-(\ref{eq:confl210}):
\begin{eqnarray}
\label{eq:BHinv11}
& & F_{01}=F_{10}\Leftrightarrow ln x \frac{B_{01}}{B_{31}} + \frac{C_1}{C_3}
= \frac{B_{01}}{B_{41} + B_{42} (ln x + 1)} + \frac{C_0}{C_4}\Rightarrow B_{12}=0,\ \frac{C_1}{C_3}=\frac{C_0}{C_4};\nonumber\\
& & F_{02}=F_{20}\Leftrightarrow ln x \frac{B_{22}}{B_{31}} + \frac{C_2}{C_3} = \frac{B_{01}}{B_{51} + B_{52} (ln x + 1)} + \frac{C_0}{C_5}\Rightarrow B_{22}=0,\ \frac{C_2}{C_3}=\frac{C_0}{C_5};\nonumber\\
& & F_{12}=F_{21}\Leftrightarrow \frac{B_{22}}{B_{42}} + \frac{C_2}{C_4} = \frac{B_{11}}{B_{51} + B_{52} (ln x + 1)} + \frac{C_1}{C_5}\Rightarrow\frac{C_2}{C_4}=\frac{C_1}{C_5}.
\end{eqnarray}
In (\ref{eq:BHinv11}), the constants $A_i, B_{ij},C_k$ are related to the constants $A_i, B_{ij}, C_k$
via matrix elements of $M$ of (\ref{eq:PFbasis2}). Therefore, one gets the following form of $F_{IJ}$:
\begin{equation}
\label{eq:BHinv12}
F_{IJ}=\left(\begin{array}{ccc} \frac{B_{01}}{C_3} + \frac{C_0}{C_3} & \frac{C_1}{C_3} & \frac{C_2}{C_3}\\
\frac{C_1}{C_3} & \frac{C_1}{C_4} & \frac{C_2}{C_4}\\
\frac{C_2}{C_3} & \frac{C_2}{C_4} & \frac{C_2}{C_5}
\end{array}\right)
\end{equation}
Using (\ref{eq:BHinv12}), one can evaluate $X^I Im(F_{IJ}) X^J$ - this is done in appendix D.

Using (\ref{eq:BHinv12}), (\ref{eq:BHinv13}) - (\ref{eq:BHinv14}), one gets:
\begin{eqnarray}
\label{eq:BHinv15}
& & {\cal N}=\nonumber\\
& & \hskip -2cm\left(\begin{array}{ccc}
a_{00}+b^{(1)}_{00}x + b^{(2)}_{00} x ln x + c_{00}(\rho-\rho_0) &
a_{01}+b^{(1)}_{01}x + b^{(2)}_{01} x ln x + c_{01}(\rho-\rho_0) &
a_{02}+b^{(1)}_{02}x + b^{(2)}_{02} x ln x + c_{02}(\rho-\rho_0) \\
a_{01}+b^{(1)}_{01}x + b^{(2)}_{01} x ln x + c_{01}(\rho-\rho_0) &
a_{11}+b^{(1)}_{11}x + b^{(2)}_{11} x ln x + c_{11}(\rho-\rho_0) &
a_{12}+b^{(1)}_{12}x + b^{(2)}_{12} x ln x + c_{12}(\rho-\rho_0) \\
a_{02}+b^{(1)}_{02}x + b^{(2)}_{02} x ln x + c_{02}(\rho-\rho_0) &
a_{12}+b^{(1)}_{12}x + b^{(2)}_{12} x ln x + c_{12}(\rho-\rho_0) &
a_{22}+b^{(1)}_{22}x + b^{(2)}_{22} x ln x + c_{22}(\rho-\rho_0)
\end{array}\right).\nonumber\\
& &
\end{eqnarray}
The constants $a_{ij}, b^{(1),(2)}_{jk},c_{lm}$ are constrained by relations, e.g.,
\begin{equation}
F_I={\cal N}_{IJ}X^J,
\end{equation}
which, e.g., for $I=0$ would imply:
\begin{eqnarray}
\label{eq:BHinv16}
& & a_{00} A_3 + a_{01} A_4 + a_{02} A_5 = A_0\nonumber\\
& & a_{00} B_{31} + b^{(1)}_{00} A_3 + a_{01} B_{41} + A_4 b^{(1)}_{01} + a_{02} B_{51} + b^{(1)}_{02} A_5
= B_{01}\nonumber\\
& & b^{(2)}_{00} A_3 + a_{01} B_{42} + b^{(2)}_{01} A_4 + a_{02} B_{52} + A_5 b^{(2)}_{02} = 0\nonumber\\
& & a_{00} C_3 + c_{00} A_3 + a_{01} C_4 + c_{01} A_4 + a_{02} C_5 + c_{02} A_5 = C_0.
\end{eqnarray}

So, substituting (\ref{eq:BHinv15}) into the last two equations of (\ref{eq:BHinv10}), one gets:
\begin{eqnarray}
\label{eq:BHinv17}
& & \partial_x({\cal P}^I{\cal N}_{IJ}{\cal P}^J)=0\Rightarrow ln x\left[({\cal P}^0)^2b^{(2)}_{00} + ({\cal P}^1)^2b^{(2)}_{11} + ({\cal P}^2)^2b^{(2)}_{22} + 2{\cal P}^0{\cal P}^1b^{(2)}_{01} + 2{\cal P}^0{\cal P}^2b^{(2)}_{02} + 2{\cal P}^1{\cal P}^2b^{(2)}_{12}\right]=0;\nonumber\\
& & \partial_{\rho-\rho_0}({\cal P}^I{\cal N}_{IJ}{\cal P}^J)=0\Rightarrow({\cal P}^0)^2c^{(2)}_{00} + ({\cal P}^1)^2c_{11} + ({\cal P}^2)^2c_{22} + 2{\cal P}^0{\cal P}^1c_{01} + 2{\cal P}^0{\cal P}^2c_{02} + 2{\cal P}^1{\cal P}^2c_{12}=0,
\end{eqnarray}
and ${\bar X}^I Im({\cal N}_{IJ}){\cal P}^J=-iw$ implies:
\begin{eqnarray}
\label{eq:BHinv18}
& & {\bar A}_I(a_{IJ}-{\bar a}_{IJ}){\cal P}^J+{\bar x}[{\bar B}_{I1}(a_{IJ} - {\bar a}_{IJ}){\cal P}^J
- {\bar b^{(1)}}_{IJ}{\bar A}_I{\cal P}^J] + x[b^{(1)}_{IJ}{\bar A}_I{\cal P}^J]
+ x ln x[{\bar A}_I b^{(2)}_{IJ}{\cal P}^J] + (\rho-\rho_0)[{\bar A}_I c_{IJ}{\cal P}^J] \nonumber\\
& & + ({\bar\rho} - {\bar\rho_0})[{\bar C}_I(a_{IJ} - {\bar a}_{IJ}){\cal P}^J - {\bar c}_{IJ}A_I{\cal P}^J]
+ {\bar x} ln{\bar x}[B_{I2}a_{IJ}{\cal P}^J]=-2{\bar w}\nonumber\\
& & {\rm or}\nonumber\\
& & \sum_{I=0}^2\Upsilon^I(x,{\bar x}, x ln x, {\bar x} ln {\bar x};\rho-\rho_0,{\bar\rho}-{\bar\rho_0}){\cal P}^I={\bar w}.
\end{eqnarray}
Using (\ref{eq:BHinv18}), we eliminate ${\cal P}^2$ from (\ref{eq:BHinv17}) to get:
\begin{eqnarray}
\label{eq:BHinv19}
& & \alpha_1({\cal P}^0)^2 + \beta_1({\cal P}^1)^2 + \gamma_1{\cal P}^0{\cal P}^1 = \lambda_1,\nonumber\\
& & \alpha_2({\cal P}^0)^2 + \beta_2({\cal P}^1)^2 + \gamma_2{\cal P}^0{\cal P}^1 = \lambda_2.
\end{eqnarray}
The equations (\ref{eq:BHinv19}) can be solved and yield four solutions which are:
\begin{eqnarray}
\label{eq:BHinv20}
& &
{\cal P}^0=
\frac{1}{2\,{\sqrt{2}}\,\Biggl( {\alpha_2}\,{\lambda_1} - {\alpha_1}\,{\lambda_2} \Biggr) }\Biggl( {\gamma_2}\,{\lambda_1} - {\gamma_1}\,{\lambda_2} +
           \sqrt{Y} \Biggr)\sqrt{X} \,\nonumber\\
& &    {{\cal P}^1}=-\frac{\sqrt{X}}{\sqrt{2}};\nonumber\\
& & {\cal P}^0=
-\frac{1}{2\,{\sqrt{2}}\,\Biggl( {\alpha_2}\,{\lambda_1} - {\alpha_1}\,{\lambda_2} \Biggr) }\Biggl( {\gamma_2}\,{\lambda_1} - {\gamma_1}\,{\lambda_2} +
           \sqrt{Y} \Biggr)\sqrt{X} \,\nonumber\\
& &    {{\cal P}^1}=\frac{\sqrt{X}}{\sqrt{2}};\nonumber\\
& & {\cal P}^0=
\frac{1}{2\,{\sqrt{2}}\,\Biggl( {\alpha_2}\,{\lambda_1} - {\alpha_1}\,{\lambda_2} \Biggr) }\Biggl( {\gamma_2}\,{\lambda_1} - {\gamma_1}\,{\lambda_2} -
           \sqrt{Y} \Biggr)\sqrt{X} \,\nonumber\\
& &    {{\cal P}^1}=-\frac{\sqrt{X}}{\sqrt{2}};\nonumber\\
& & {\cal P}^0=
-\frac{1}{2\,{\sqrt{2}}\,\Biggl( {\alpha_2}\,{\lambda_1} - {\alpha_1}\,{\lambda_2} \Biggr) }\Biggl( {\gamma_2}\,{\lambda_1} - {\gamma_1}\,{\lambda_2} -
           \sqrt{Y} \Biggr)\sqrt{X} \,\nonumber\\
& &    {{\cal P}^1}=\frac{\sqrt{X}}{\sqrt{2}}
                         \end{eqnarray}
                         where
                         \begin{eqnarray}
                         \label{eq:BHinv21}
                         & & X\equiv\frac{1}{{{\alpha_2}}^2\,{{\beta_1}}^2 +
               {\alpha_2}\,\Biggl[ -2\,{\alpha_1}\,{\beta_1}\,{\beta_2} +
                  {\gamma_1}\,\Biggl( {\beta_2}\,{\gamma_1} - {\beta_1}\,{\gamma_2} \Biggr)  \Biggr]  +
               {\alpha_1}\,\Biggl[ {\alpha_1}\,{{\beta_2}}^2 +
                  {\gamma_2}\,\Biggl( -{\beta_2}\,{\gamma_1}   + {\beta_1}\,{\gamma_2} \Biggr)
                  \Biggr] }\nonumber\\
                  & & \times\Biggl[2\,{{\alpha_2}}^2\,{\beta_1}\,{\lambda_1} +
               {\alpha_1}\,\Biggl( {{\gamma_2}}^2\,{\lambda_1} + 2\,{\alpha_1}\,{\beta_2}\,{\lambda_2} -
                  {\gamma_2}\,\Biggl( {\gamma_1}\,{\lambda_2} +
                     {\sqrt{X_1}}
                     \Biggr)  \Biggr)\Biggr],\nonumber\\
                     & & Y\equiv{{\gamma_2}}^2\,{{\lambda_1}}^2 - 2\,{\gamma_1}\,{\gamma_2}\,{\lambda_1}\,{\lambda_2} +
               4\,{\alpha_2}\,{\lambda_1}\,\Biggl( -{\beta_2}\,{\lambda_1}  +
                  {\beta_1}\,{\lambda_2} \Biggr)  +
               {\lambda_2}\,\Biggl( 4\,{\alpha_1}\,{\beta_2}\,{\lambda_1} -
                  4\,{\alpha_1}\,{\beta_1}\,{\lambda_2} + {{\gamma_1}}^2\,{\lambda_2} \Biggr),  \end{eqnarray}
                         and
                         \begin{eqnarray}
                         \label{eq:BHinv22}
                         & & X_1\equiv Y + {\alpha_2}\,
                \Biggl[ -2\,{\alpha_1}\,\Biggl( {\beta_2}\,{\lambda_1} + {\beta_1}\,{\lambda_2} \Biggr)  +
                  {\gamma_1}\,\Biggl( -{\gamma_2}\,{\lambda_1}  + {\gamma_1}\,{\lambda_2}\nonumber\\
                  & &  +
                     \sqrt{{{\gamma_2}}^2\,{{\lambda_1}}^2 -
                         2\,{\gamma_1}\,{\gamma_2}\,{\lambda_1}\,{\lambda_2} +
                         4\,{\alpha_2}\,{\lambda_1}\,
                          \Biggl( -{\beta_2}\,{\lambda_1}   + {\beta_1}\,{\lambda_2} \Biggr)  +
                         {\lambda_2}\,\Biggl( 4\,{\alpha_1}\,{\beta_2}\,{\lambda_1} -
                            4\,{\alpha_1}\,{\beta_1}\,{\lambda_2} + {{\gamma_1}}^2\,{\lambda_2} \Biggr)}\ \Biggr)\Biggr].\nonumber\\
                            & &
                            \end{eqnarray}
 One can show that one does get ${\cal P}^I\sim X^I$ as one of the solutions - this corresponds to a supersymmetric
black hole, and the other solutions correspond to non-supersymmetric black holes.

\section{``Fake Superpotentials"}

In this section, using the results of \cite{Ceresole+Dall'agata}, we show the existence of ``fake superpotentials" corresponding to black-hole solutions
for type II compactification on (\ref{eq:hypersurface}).

As argued in \cite{Ceresole+Dall'agata}, dS-curved domain wall solutions in gauged supergravity and non-extremal black hole solutions in Maxwell-Einstein theory  have the same effective action. In the context of domain wall solutions, if there exists a ${\cal W}(z^i,{\bar z}^i)\in{\bf R}: V_{DW}(\equiv{\rm Domain\ Wall\ Potential})=-{\cal W}^2 + \frac{4}{3\gamma^2}g^{i{\bar j}}\partial_i{\cal W}\partial_{\bar j}{\cal W}$, $z^i$ being complex scalar fields, then the solution to the second-order equations for domain walls, can also be derived from the following first-order flow equations: $U^\prime=\pm e^U\gamma(r){\cal W};\ (z^i)^\prime = \mp e^U\frac{2}{\gamma^2}g^{i{\bar j}}\partial_{\bar j}{\cal W}$, where
$\gamma\equiv\sqrt{1 + \frac{e^{-2U}\Lambda}{{\cal W}^2}}$.

Now, spherically symmetric, charged, static and asymptotically flat black hole solutions of Einstein-Maxwell theory coupled to complex scalar fields have the form: $dz^2 = - e^{2U(r)} dt^2 + e^{-2U(r)}\biggl[\frac{c^4}{sinh^4(cr)} dr^2 $ $+ \frac{c^2}{sinh^2(cr)}(d\theta^2 + sin^2\theta d\phi^2)\biggr]$, where the non-extremality parameter
$c$ gets related to the positive cosmological constant $\Lambda>0$ for domain walls. For non-constant scalar fields, only for $c=0$ that corresponds to extremal black holes, one can write down first-order flow equations in terms of a ${\cal W}(z^i,{\bar z}^i)\in{\bf R}$: $U^\prime=\pm e^U{\cal W};\ (z^i)^\prime=\pm2e^U g^{i{\bar j}}\partial_{\bar j}{\cal W},$ and the potential $\tilde{V}_{BH}\equiv {\cal W}^2 + 4g^{i{\bar j}}\partial_i{\cal W}\partial_{\bar j}{\cal W}$ can be compared with the ${\cal N}=2$ supergravity black-hole potential $V_{BH}=|Z|^2+g^{i{\bar j}}D_iZD_{\bar j}{\bar Z}$ by identifying
${\cal W}\equiv|Z|$. For non-supersymmetric theories or supersymmetric theories where the black-hole constraint equation admits multiple solutions which may happen because several ${\cal W}$s may correspond to the same $\tilde{V}_{BH}$ of which only one choice of ${\cal W}$ would correspond to the true central charge, one hence talks about ``fake superpotential" or ``fake supersymmetry" - a ${\cal W}:\partial_i{\cal W}=0$ would correspond to a stable non-BPS black hole. Defining ${\cal V}\equiv e^{2U}V(z^i,{\bar z}^i),
{\cal\bf W}\equiv e^U{\cal W}(z^i,{\bar z}^i)$, one sees that ${\cal V}(x^A\equiv U,z^i,{\bar z}^i)=g^{AB}\partial_A{\bf W}(x)\partial_B{\bf W}(x)$, where $g_{UU}=1$ and $g_{Ui}=0$. This illustrates the fact that one gets the same potential ${\cal V}(x)$ for all vectors $\partial_A{\cal\bf W}$
with the same norm. In other words, ${\bf W}$ and $\tilde{\bf W}$ defined via: $\partial_A{\bf W}=R_A^{\ B}(z,{\bar z})\partial_B\tilde{\bf W}$ correspond to the same ${\cal V}$ provided:
$R^TgR=g$.

For ${\cal N}=2$ supergravity, the black hole potential $V_{BH}=Q^T{\cal M}Q$ where $Q=(p^\Lambda,q_\Lambda)$ is an $Sp(2n_v+2,{\bf Z})$-valued vector ($n_V$ being the number of vector multiplets) and ${\cal M}\in Sp(2n_V+2)$ is given by:
\begin{equation}
\label{eq:FakeW1}
{\cal M}=\left(\begin{array}{cc}
A&B\\
C&D
\end{array}\right),
\end{equation}
where
\begin{eqnarray}
\label{eq:FakeW2}
& & A\equiv Re {\cal N} (Im {\cal N})^{-1}\nonumber\\
& & B\equiv-Im {\cal N} - Re {\cal N} (Im {\cal N})^{-1} Re {\cal N} \nonumber\\
& & C\equiv (Im {\cal N})^{-1} \nonumber\\
& & D = -A^T = - (Im {\cal N}^{-1})^T (Re {\cal N})^T.
\end{eqnarray}
Defining $M: {\cal M}={\cal I}M$ where
\begin{eqnarray}
\label{eq:FakeW3}
& & M\equiv\left(\begin{array}{cc}
D&C\\
B&A
\end{array}\right),\nonumber\\
& & {\cal I}\equiv\left(\begin{array}{cc}
0&-{\bf 1}_{n_V+1}\\
{\bf 1}_{n_V+1}&0
\end{array}\right).
\end{eqnarray}
The central charge $Z=e^{\frac{K}{2}}(q_\Lambda X^\Lambda - p^\Lambda F_\lambda)$, a symplectic invariant
is expressed as a symplectic dot product of $Q$ and covariantly holomorphic sections: ${\cal V}\equiv e^{\frac{K}{2}}(X^\Lambda,F_\Lambda)=(L^\Lambda,M_\Lambda) (M_\Lambda={\cal N}_{\Lambda\Sigma}L^\Sigma)$,
and hence can be written as
\begin{equation}
\label{eq:FakeW4}
Z = Q^T{\cal I}{\cal V} = L^\Lambda q_\Lambda - M_\lambda p^\Lambda.
\end{equation}
Now, the black-hole potential $V_{BH}=Q^T{\cal M}Q$ (being a symplectic invariant) is invariant under:
\begin{eqnarray}
\label{eq:FakeW5}
& & Q\rightarrow SQ,\nonumber\\
& & S^T{\cal M}S={\cal M}.
\end{eqnarray}
As $S$ is a symplectic matrix, $S^T{\cal I}={\cal I}S^{-1}$, which when substituted in (\ref{eq:FakeW5})
yields:
\begin{equation}
\label{eq:fakeW6}
[S,M]=0.
\end{equation}
In other words, if there exists a constant symplectic matrix $S:[S,M]=0$, then there exists a fake superpotential $Q^TS^T{\cal I}{\cal V}$ whose critical points, if they exist, describe non-supersymmetric
black holes.

We now construct an explicit form of $S$. For concreteness, we work at the point in the moduli space for
(\ref{eq:hypersurface}): $\phi^3=1$ and large $\psi$ near $x=0$ and $\rho=\rho_0$. Given the form of ${\cal N}_{IJ}$ in (\ref{eq:BHinv17}), one sees that:
\begin{eqnarray}
\label{eq:FakeW7}
& & {\cal N}^{-1}=\nonumber\\
& & \hskip -2cm\left(\begin{array}{ccc}
\tilde{a}_{00}+\tilde{b}^{(1)}_{00}x + \tilde{b}^{(2)}_{00} x ln x + \tilde{c}_{00}(\rho-\rho_0) &
\tilde{a}_{01}+\tilde{b}^{(1)}_{01}x + \tilde{b}^{(2)}_{01} x ln x + \tilde{c}_{01}(\rho-\rho_0) &
\tilde{a}_{02}+\tilde{b}^{(1)}_{02}x + \tilde{b}^{(2)}_{02} x ln x + \tilde{c}_{02}(\rho-\rho_0) \\
\tilde{a}_{01}+\tilde{b}^{(1)}_{01}x + \tilde{b}^{(2)}_{01} x ln x + \tilde{c}_{01}(\rho-\rho_0) &
\tilde{a}_{11}+\tilde{b}^{(1)}_{11}x + \tilde{b}^{(2)}_{11} x ln x + \tilde{c}_{11}(\rho-\rho_0) &
\tilde{a}_{12}+\tilde{b}^{(1)}_{12}x + \tilde{b}^{(2)}_{12} x ln x + \tilde{c}_{12}(\rho-\rho_0) \\
\tilde{a}_{02}+\tilde{b}^{(1)}_{02}x + \tilde{b}^{(2)}_{02} x ln x + \tilde{c}_{02}(\rho-\rho_0) &
\tilde{a}_{12}+\tilde{b}^{(1)}_{12}x + \tilde{b}^{(2)}_{12} x ln x + \tilde{c}_{12}(\rho-\rho_0) &
\tilde{a}_{22}+\tilde{b}^{(1)}_{22}x + \tilde{b}^{(2)}_{22} x ln x + \tilde{c}_{22}(\rho-\rho_0)
\end{array}\right),\nonumber\\
& &
\end{eqnarray}
which as expected is symmetric (and hence so will $Re {\cal N}$ and $(Im {\cal N})^{-1}$). One can therefore write
\begin{equation}
\label{eq:fakeW8}
M\equiv\left(\begin{array}{cc}
U & V \\
X & -U^T
\end{array}\right),
\end{equation}
where $V^T=V,\ X^T=X$ and $U, V, X$ are $3\times 3$ matrices constructed from $Re {\cal N}$ and
$(Im {\cal N})^{-1}$. Writing
\begin{equation}
\label{eq:FakeW9}
S=\left(\begin{array}{cc}
{\cal A} & {\cal B} \\
{\cal C} & {\cal D}
\end{array}\right),
\end{equation}
(${\cal A}, {\cal B}, {\cal C}, {\cal D}$ are $3\times3$ matrices) and given that $S\in Sp(6)$, implying:
\begin{equation}
\label{eq:FakeW10}
\left(\begin{array}{cc}
{\cal A}^T & {\cal C}^T\\
{\cal B}^T & {\cal D}^T
\end{array}\right)\left(\begin{array}{cc}
0 & -{\bf 1}_3 \\
{\bf 1}_3 & 0
\end{array}\right)\left(\begin{array}{cc}
{\cal A} & {\cal B} \\
{\cal C} & {\cal D}
\end{array}\right)=\left(\begin{array}{cc}
0 & -{\bf 1}_3 \\
{\bf 1}_3 & 0
\end{array}\right),
\end{equation}
which in turn implies the following matrix equations:
\begin{eqnarray}
\label{eq:FakeW11}
& & -{\cal A}^T{\cal C} + {\cal C}^T{\cal A} = 0,\nonumber\\
& & -{\cal B}^T{\cal D} + {\cal D}^T{\cal B} = 0,\nonumber\\
& & -{\cal A}^T{\cal D} + {\cal C}^T{\cal B} = - {\bf 1}_3,\nonumber\\
& & -{\cal B}^T{\cal C} + {\cal D}^T{\cal A} = {\bf 1}_3.
\end{eqnarray}
Now, $[S,M]=0$ implies:
\begin{equation}
\label{eq:FakeW12}
\left(\begin{array}{cc}
{\cal A} U + {\cal B} X & {\cal A} V - {\cal B} U^T \\
{\cal C} U + {\cal D} X & {\cal C} V - {\cal D} U^T
\end{array}\right) = \left(\begin{array}{cc}
U {\cal A} + V {\cal C} & U {\cal B} + V {\cal D} \\
X {\cal A} - U^T {\cal C} & X {\cal B} - U^T {\cal D}
\end{array}\right).
\end{equation}
The system of equations (\ref{eq:FakeW11}) can be satisfied, e.g., by the following choice of ${\cal A}, {\cal B}, {\cal C}, {\cal D}$:
\begin{equation}
\label{eq:FakeW13}
{\cal B} = {\cal C}=0;\ {\cal D} = ({\cal A}^{-1})^T.
\end{equation}
To simplify matters further, let us assume that ${\cal A}\in O(3)$ implying that $({\cal A}^{-1})^T = {\cal A}$. Then (\ref{eq:FakeW12}) would imply:
\begin{eqnarray}
\label{eq:FakeW14}
& & [{\cal A},V] = 0,\nonumber\\
& & [{\cal A},X] = 0,\nonumber\\
& & [{\cal A}^{-1},U] = 0,\nonumber\\
& & [{\cal A},U] = 0.
\end{eqnarray}
For points near the conifold locus $\phi=\omega^{-1},\rho=\rho_0$, using (\ref{eq:confl24})-(\ref{eq:confl210}) and (\ref{eq:BHinv12}) and dropping the moduli-dependent terms in (\ref{eq:Wconfl11}), one can show:
\begin{eqnarray}
\label{eq:FakeW141}
& & \left(Im {\cal N}^{-1}\right)_{0I}\left(Re {\cal N}\right)_{IK}=0,\ K=1,2\nonumber\\
& & \left(Im {\cal N}^{-1}\right)_{0K}=0,\ K=1,2\nonumber\\
& & \left(Im {\cal N}\right)_{0K} + \left(Re {\cal N}\right)_{0I}\left(Im {\cal N}^{-1}\right)_{IJ}
\left(Re{\cal N}\right)_{JK} = 0,\ K=1,2.
\end{eqnarray}
This is equivalent to saying that the first two and the last equations in (\ref{eq:FakeW14}) can be satisfied
by:
\begin{equation}
\label{eq:FakeW15}
{\cal A} = \left(\begin{array}{ccc}
1&0&0\\
0&-1&0\\
0&0&-1
\end{array}\right).
\end{equation}
The form of $A$ chosen in (\ref{eq:FakeW15}) also satisfies the third equation in (\ref{eq:FakeW14}) - similar solutions were also
considered in \cite{Ceresole+Dall'agata}. Hence,
\begin{equation}
\label{eq:FakeW16}
S = \left(\begin{array}{cccccc}
1&0&0&0&0&0\\
0&-1&0&0&0&0\\
0&0&-1&0&0&0\\
0&0&0&1&0&0\\
0&0&0&0&-1&0\\
0&0&0&0&0&-1
\end{array}\right).
\end{equation}
We therefore see that the non-supersymmetric black-hole corresponding to the fake superpotential
$Q^TS^T{\cal I}{\cal V}$, $S$ being given by (\ref{eq:FakeW16}), corresponds to the change of sign of
two of the three electric and magetic charges as compared to a supersymmetric black hole. The symmetry
properties of the elements of ${\cal M}$ and hence $M$ may make it generically possible to find a constant
$S$ like the one in (\ref{eq:FakeW16}) for two-paramater Calabi-Yau compactifications.

\section{Conclusion}

We looked at several aspects of complex structure moduli stabilization and inclusion, in the large volume limit,
of perturbative and specially non-perturbative $\alpha^\prime$-corrections and instanton contributions in the
K\"{a}hler potential and superpotential in the context of K\"{a}hler moduli, for a two-parameter ``Swiss cheese" Calabi-Yau three-fold of a projective variety expressed as a (resolution of a) hypersurface in a complex weighted
projective space, with mutliple conifold loci in its moduli space. As regards ${\cal N}=1$
type IIB compactifications on orientifold of the aforementioned Calabi-Yau, we argued the existence of
(extended) ``area codes" wherein for the same values of the RR and NS-NS fluxes, one is able to stabilize
the complex structure and axion-dilaton moduli at points away from and close to the two singular conifold
loci. It would be nice to explicitly work out the numerics and find the set of fluxes
corresponding to the aforementioned area codes (whose existence we argued), as well as the flow of the
moduli corresponding to the domain walls arising as a consequence of
such area codes. Further, in the large volume limit of the orientifold, we show that with the inclusion of
non-perturbative $\alpha^\prime$-corrections that survive the orientifolding alongwith the nonperturbative
contributions from instantons, it is possible to get a non-supersymmetric dS minimum
{\it without the inclusion of anti-D3 branes}.
It would interesting to investigate the effect of string loop
corrections in the context of orientifolds of compact Calabi-Yau of the type considered in this work (See \cite{BHP}).
As regards supersymmetric and non-supersymmetric black-hole attractors
in ${\cal N}=2$ type II compactifications on the same Calabi-Yau three-fold, we explicitly solve the
``inverse problem" of determining the electric and magnetic charges of an extremal black hole given the
extremum values of the moduli. In the same context, we also show explicitly the existence of ``fake superpotentials"
as a consequence of non-unique superpotentials for the same black-hole potential corresponding to reversal
of signs of some of  the electric and magnetic charges. There may be interesting connection between the existence
of such fake superpotentials and works like \cite{EH}\footnote{AM thanks S.Mathur for bringing \cite{EH} to
our attention.}

\section*{Acknowledgements}

PS is supported by a C.S.I.R., Government of India, (junior) research fellowship and the work of AM is partially
supported by D.A.E. (Government of India) Young Scientist Award project grant. AM also thanks the CERN
theory group for hosptiality and the theoretical high energy physics groups at
McGill University  (specially Keshav Dasgupta),
University of Pennsylvania (specially Vijay Balasubramanian), Ohio State University (specially
Samir Mathur) and Columbia University (specially D.Kabat), for hospitality where part of this work was done and also where preliminary versions of this
work were presented. AM also thanks Samir Mathur and Jeremy Michelson for interesting discussions, and Rajesh Gopakumar and Ashoke Sen for interesting questions (that helped improve the arguments of section 4) during AM's presentation of part of the material of this paper at ISM07, Harish-Chandra Research Institute, Allahabad.

\appendix
\section{Periods}
\setcounter{equation}{0}
\seceqaa

In this appendix, we fill in the details relevant to evaluation of periods in different portions of the complex structure moduli space of section ${\bf 2}$.

$\underline{|\phi^3|<1,\ {\rm large}\ \psi}$

The expressions for $P_{1,2,3}$ relevant to (\ref{eq:smallphilargepsi2}) in section {\bf 2} are given as under:

\begin{equation}
\label{eq:smallphilargepsi3}
P_1\equiv\left(\begin{array}{c}
\sum_{m=0}^\infty\sum_{n=0}^\infty A_{m,n}\frac{\phi_0^m}{\rho_0^{18n + 6m}}\\
\sum_{m=0}^\infty\sum_{n=0}^\infty e^{\frac{i\pi(-35m + 128 n)}{9}}
A_{m,n}\frac{\phi_0^m}{\rho_0^{18n + 6m}}\\
\sum_{m=0}^\infty\sum_{n=0}^\infty e^{\frac{2i\pi(-35m + 128 n)}{9}}
A_{m,n}\frac{\phi_0^m}{\rho_0^{18n + 6m}}\\
\sum_{m=0}^\infty\sum_{n=0}^\infty e^{\frac{i\pi(-35m + 128 n)}{3}}
A_{m,n}\frac{\phi_0^m}{\rho_0^{18n + 6m}}\\
\sum_{m=0}^\infty\sum_{n=0}^\infty e^{\frac{4i\pi(-35m + 128 n)}{9}}
A_{m,n}\frac{\phi_0^m}{\rho_0^{18n + 6m}}\\
\sum_{m=0}^\infty\sum_{n=0}^\infty e^{\frac{5i\pi(-35m + 128 n)}{9}}
A_{m,n}\frac{\phi_0^m}{\rho_0^{18n + 6m}}\\
\end{array}\right),
\end{equation}
\begin{equation}
\label{eq:smallphilargepsi4}
P_2\equiv\left(\begin{array}{c}
\sum_{m=0}^\infty\sum_{n=0}^\infty A_{m,n}\frac{m\phi_0^{m-1}}{\rho_0^{18n + 6m}}\\
\sum_{m=0}^\infty\sum_{n=0}^\infty A_{m,n}\frac{m\phi_0^{m-1}}{\rho_0^{18n + 6m}}
e^{\frac{i\pi(-35m + 128 n)}{9}}\\
\sum_{m=0}^\infty\sum_{n=0}^\infty A_{m,n}\frac{m\phi_0^{m-1}}{\rho_0^{18n + 6m}}
e^{2\frac{i\pi(-35m + 128 n)}{9}}\\
\sum_{m=0}^\infty\sum_{n=0}^\infty A_{m,n}\frac{m\phi_0^{m-1}}{\rho_0^{18n + 6m}}
e^{\frac{i\pi(-35m + 128 n)}{3}}\\
\sum_{m=0}^\infty\sum_{n=0}^\infty A_{m,n}\frac{m\phi_0^{m-1}}{\rho_0^{18n + 6m}}
e^{\frac{4i\pi(-35m + 128 n)}{9}}\\
\sum_{m=0}^\infty\sum_{n=0}^\infty A_{m,n}\frac{m\phi_0^{m-1}}{\rho_0^{18n + 6m}}
e^{\frac{5i\pi(-35m + 128 n)}{9}}\\
\end{array}\right)
\end{equation}
and
\begin{equation}
\label{eq:smallphilargepsi5}
P_3\equiv\left(\begin{array}{c}
-\sum_{m=0}^\infty\sum_{n=0}^\infty A_{m,n}\frac{(18n + 6m)\phi_0^m}{\rho_0^{18n + 6m + 1}}\\
-\sum_{m=0}^\infty\sum_{n=0}^\infty A_{m,n}\frac{(18n + 6m)\phi_0^m}{\rho_0^{18n + 6m + 1}}e^{\frac{i\pi(-35m + 128 n)}{9}}\\
-\sum_{m=0}^\infty\sum_{n=0}^\infty A_{m,n}\frac{(18n + 6m)\phi_0^m}{\rho_0^{18n + 6m + 1}}e^{\frac{2i\pi(-35m + 128 n)}{9}}\\
-\sum_{m=0}^\infty\sum_{n=0}^\infty A_{m,n}\frac{(18n + 6m)\phi_0^m}{\rho_0^{18n + 6m + 1}}e^{\frac{i\pi(-35m + 128 n)}{3}}\\
-\sum_{m=0}^\infty\sum_{n=0}^\infty A_{m,n}\frac{(18n + 6m)\phi_0^m}{\rho_0^{18n + 6m + 1}}e^{\frac{4i\pi(-35m + 128 n)}{9}}\\
-\sum_{m=0}^\infty\sum_{n=0}^\infty A_{m,n}\frac{(18n + 6m)\phi_0^m}{\rho_0^{18n + 6m + 1}}e^{\frac{5i\pi(-35m + 128 n)}{9}}\\
\end{array}\right).
\end{equation}
The coefficients $A_{m,n}$ appearing in (\ref{eq:smallphilargepsi3})-(\ref{eq:smallphilargepsi5}) are given by:
$$A_{m,n}\equiv \frac{(18n + 6m)!(-3\phi)^m(3^4.2)^{18n + 6m}}{(9n + 3m)!(6n + 2m)!(n!)^3m!18^{18n + 6m}}.$$
The equations (\ref{eq:smallphilargepsi3})-(\ref{eq:smallphilargepsi5}) will be used in obtaining (\ref{eq:Wawayconfl}) and the third set of equations in (\ref{eq:DW=0}).

$\underline{|\frac{\rho^6}{\phi - \omega^{0,-1,-2}}|<1}$

The expressions for $M_{1,2,3}$ relevant to (\ref{eq:awaycl12}) in section are given as under:

\begin{equation}
\label{eq:awaycl13}
M_1\equiv\left(\begin{array}{c}
\sum_{r=1,5}\sum_{k=0}^\infty\sum_{m=0}^\infty A_{k,m,r}\rho_0^{6k+r}\phi_0^m
\\
\sum_{r=1,5}\sum_{k=0}^\infty\sum_{m=0}^\infty A_{k,m,r}\rho_0^{6k+r}\phi_0^m
e^{\frac{2i\pi(k+\frac{r}{6})}{3} + \frac{2i\pi m}{3}}
\\
\sum_{r=1,5}\sum_{k=0}^\infty\sum_{m=0}^\infty A_{k,m,r}\rho_0^{6k+r}\phi_0^m
 e^{\frac{4i\pi(k+\frac{r}{6})}{3} + \frac{4i\pi m}{3}}
 \\
\sum_{r=1,5}\sum_{k=0}^\infty\sum_{m=0}^\infty A_{k,m,r}\rho_0^{6k+r}\phi_0^m e^{\frac{i\pi r}{3}}
\\
\sum_{r=1,5}\sum_{k=0}^\infty\sum_{m=0}^\infty A_{k,m,r}\rho_0^{6k+r}\phi_0^m
e^{\frac{2i\pi(k+\frac{r}{6})}{3} + \frac{2i\pi m}{3} + \frac{i\pi r}{3}}
\\
\sum_{r=1,5}\sum_{k=0}^\infty\sum_{m=0}^\infty A_{k,m,r}\rho_0^{6k+r}\phi_0^m
e^{\frac{4i\pi(k+\frac{r}{6})}{3} + \frac{4i\pi m}{3} + \frac{i\pi r}{3}}
\end{array}\right),
\end{equation}
\begin{equation}
\label{eq:awaycfl4}
M_2\equiv\left(\begin{array}{c}
\sum_{r=1,5}\sum_{k=0}^\infty\sum_{m=0}^\infty A_{k,m,r}(6k + r)\rho_0^{6k + r - 1}\phi_0^m
\\
\sum_{r=1,5}\sum_{k=0}^\infty\sum_{m=0}^\infty A_{k,m,r}(6k + r)\rho_0^{6k + r - 1}\phi_0^m
e^{\frac{2i\pi(k+\frac{r}{6})}{3} + \frac{2i\pi m}{3}}
\\
\sum_{r=1,5}\sum_{k=0}^\infty\sum_{m=0}^\infty A_{k,m,r}(6k + r)\rho_0^{6k + r - 1}\phi_0^m
 e^{\frac{4i\pi(k+\frac{r}{6})}{3} + \frac{4i\pi m}{3}}
 \\
\sum_{r=1,5}\sum_{k=0}^\infty\sum_{m=0}^\infty A_{k,m,r}(6k + r)\rho_0^{6k + r - 1}\phi_0^{m - 1} e^{\frac{i\pi r}{3}}\\
\sum_{r=1,5}\sum_{k=0}^\infty\sum_{m=0}^\infty A_{k,m,r}(6k + r)\rho_0^{6k + r - 1}\phi_0^m e^{\frac{2i\pi(k+\frac{r}{6})}{3} + \frac{2i\pi m}{3} + \frac{i\pi r}{3}}
\\
\sum_{r=1,5}\sum_{k=0}^\infty\sum_{m=0}^\infty A_{k,m,r}(6k + r)\rho_0^{6k + r - 1}\phi_0^m
e^{\frac{4i\pi(k+\frac{r}{6})}{3} + \frac{4i\pi m}{3} + \frac{i\pi r}{3}}
\end{array}\right)
\end{equation}
and
\begin{equation}
\label{eq:awaycl15}
M_3\equiv\left(\begin{array}{c}
\sum_{r=1,5}\sum_{k=0}^\infty\sum_{m=0}^\infty A_{k,m,r}m\rho_0^{6k + r}\phi_0^{m - 1}\\
\sum_{r=1,5}\sum_{k=0}^\infty\sum_{m=0}^\infty A_{k,m,r}m\rho_0^{6k + r}\phi_0^{m - 1}
e^{\frac{2i\pi(k+\frac{r}{6})}{3} + \frac{2i\pi m}{3}}\\
\sum_{r=1,5}\sum_{k=0}^\infty\sum_{m=0}^\infty A_{k,m,r}m\rho_0^{6k + r}\phi_0^{m - 1}
e^{\frac{4i\pi(k+\frac{r}{6})}{3} + \frac{4i\pi m}{3}}\\
\sum_{r=1,5}\sum_{k=0}^\infty\sum_{m=0}^\infty A_{k,m,r}m
\rho_0^{6k + r}\phi_0^{m-1} e^{\frac{i\pi r}{3}}\\
\sum_{r=1,5}\sum_{k=0}^\infty\sum_{m=0}^\infty A_{k,m,r}m
\rho_0^{6k + r}\phi_0^{m - 1} e^{\frac{2i\pi(k+\frac{r}{6})}{3} + \frac{2i\pi m}{3} + \frac{i\pi r}{3}}\\
\sum_{r=1,5}\sum_{k=0}^\infty\sum_{m=0}^\infty A_{k,m,r}m\rho_0^{6k + r}\phi_0^{m - 1}
e^{\frac{4i\pi(k+\frac{r}{6})}{3} + \frac{4i\pi m}{3} + \frac{i\pi r}{3}}
\end{array}\right).
\end{equation}
In equations (\ref{eq:awaycl13})-(\ref{eq:awaycl15}), the coefficients $A_{k,m,r}$ are given by:
$$A_{k,m,r}\equiv e^{\frac{i\pi ar}{3}}sin\left(\frac{\pi r}{3}\right)\frac{(-)^k3^{-1 + k+\frac{r}{6} + m}e^{-\frac{i\pi(k + \frac{r}{6})}{3} + \frac{2im\pi}{3}}\Gamma(\frac{m + k + \frac{r}{6}}{3})(\Gamma(k+\frac{r}{6}))^2}{(\Gamma(1 - \frac{m + k + \frac{r}{6}}{3}))^2m!}.$$ The equations (\ref{eq:awaycl13})-(\ref{eq:awaycl15}) will be used in obtaining (\ref{eq:Wawayconfl}) and the fourth set of equations in  (\ref{eq:DW=0}).

$\underline{{\rm Near\ the\ conifold\ locus:}\ \rho^6 + \phi = 1}$

The expressions for $N_{1,2,3}$ relevant for evaluation of (\ref{eq:confl14}) in section {\bf 2}, are given as under:

\begin{equation}
\label{eq:confl15}
N_1\equiv\left(\begin{array}{c}
\sum_{r=1,5}\sum_{k=0}^\infty A_{k,0,r}e^{-\frac{i\pi(k + \frac{r}{6})}{3}}\\
\sum_{r=1,5}\sum_{k=0}^\infty A_{k,0,r}e^{\frac{i\pi(k + \frac{r}{6})}{3}}\\
\sum_{r=1,5}\sum_{k=0}^\infty A_{k,0,r}e^{i\pi(k + \frac{r}{6}}\\
\sum_{r=1,5}\sum_{k=0}^\infty A_{k,0,r}e^{\frac{i\pi(-k + 5\frac{r}{6})}{3}}\\
\sum_{r=1,5}\sum_{k=0}^\infty A_{k,0,r}e^{-\frac{i\pi(k + 7\frac{r}{6})}{3}}\\
\sum_{r=1,5}\sum_{k=0}^\infty A_{k,0,r}e^{-\frac{i\pi(3k + 3\frac{r}{6})}{3}}\end{array}\right),
\end{equation}
\begin{equation}
\label{eq:confl16}
N_2\equiv\left(\begin{array}{c}
\sum_{r=1,5}\sum_{k=0}^\infty (k + \frac{r}{6}) A_{k,0,r}e^{-\frac{i\pi(k + \frac{r}{6})}{3}}\\
\sum_{r=1,5}\sum_{k=0}^\infty (k + \frac{r}{6}) A_{k,0,r}e^{\frac{i\pi(k + \frac{r}{6})}{3}}\\
\sum_{r=1,5}\sum_{k=0}^\infty (k + \frac{r}{6}) A_{k,0,r}e^{i\pi(k + \frac{r}{6}}\\
\sum_{r=1,5}\sum_{k=0}^\infty (k + \frac{r}{6}) A_{k,0,r}e^{\frac{i\pi(-k + 5\frac{r}{6})}{3}}\\
\sum_{r=1,5}\sum_{k=0}^\infty (k + \frac{r}{6}) A_{k,0,r}e^{-\frac{i\pi(k + 7\frac{r}{6})}{3}}\\
\sum_{r=1,5}\sum_{k=0}^\infty (k + \frac{r}{6}) A_{k,0,r}e^{-\frac{i\pi(3k + 3\frac{r}{6})}{3}}\end{array}\right)
\end{equation}
and
\begin{equation}
\label{eq:confl17}
N_3\equiv\left(\begin{array}{c}
\sum_{r=1,5}\sum_{k=0}^\infty (A_{k,1,r} e^{\frac{2i\pi}{3}} - A_{k,0,r}(k + \frac{r}{6}))
e^{-\frac{i\pi(k + \frac{r}{6})}{3}}\\
\sum_{r=1,5}\sum_{k=0}^\infty (A_{k,1,r} e^{\frac{4i\pi}{3}} - A_{k,0,r}(k + \frac{r}{6}))
e^{\frac{i\pi(k + \frac{r}{6})}{3}}\\
\sum_{r=1,5}\sum_{k=0}^\infty (A_{k,1,r}  - A_{k,0,r}(k + \frac{r}{6}))
e^{i\pi(k + \frac{r}{6}}\\
\sum_{r=1,5}\sum_{k=0}^\infty (A_{k,1,r} e^{\frac{2i\pi}{3}} - A_{k,0,r}(k + \frac{r}{6}))
e^{\frac{i\pi(-k + 5\frac{r}{6})}{3}}\\
\sum_{r=1,5}\sum_{k=0}^\infty (A_{k,1,r} e^{\frac{4i\pi}{3}} - A_{k,0,r}(k + \frac{r}{6}))
e^{-\frac{i\pi(k + 7\frac{r}{6})}{3}}\\
\sum_{r=1,5}\sum_{k=0}^\infty (A_{k,1,r}  - A_{k,0,r}(k + \frac{r}{6}))
e^{-\frac{i\pi(3k + 3\frac{r}{6})}{3}}
\end{array}\right).
\end{equation}
The coefficients $A_{k,m,r}$ figuring in (\ref{eq:confl15})-(\ref{eq:confl17}) are given by: 
$$A_{k,m,r}\equiv \frac{3^{-1 + k + \frac{r}{6} + m}e^{\frac{2i\pi(\sigma + 1)}{3} + \frac{-i\pi(k + \frac{r}{6})}{3}}(-)^k(\Gamma(k + \frac{r}{6}))^2}{\Gamma(k + 1)\Gamma(k + \frac{r}{6})\Gamma(k + \frac{r}{6})(\Gamma(1 - \frac{m + k + \frac{r}{6}}{3}))^2m!}sin\left(\frac{\pi r}{3}\right).$$

$\underline{{\rm Near}\ \phi^3=1,\ {\rm Large}\ \rho}$

The expressions for $\varpi_{0,...,5}$ relevant for evaluation of (\ref{eq:confl210}) in section {\bf 2}, are given as under:

\begin{eqnarray}
\label{eq:confl24}
& &
(i) \varpi_0 \sim -\int_{\Gamma^\prime}\frac{d\mu}{4\pi^2i}\frac{\Gamma(-\mu)\Gamma(\mu+\frac{1}{6})}{\Gamma(1+\mu)}\rho^{-6\mu}
\frac{\sqrt{3}\sum_{\tau=0}^2\gamma^{0,\tau}_\mu\omega^\tau(\phi-\omega^{-\tau})^{\mu+1}}{2\pi(\mu+1)}\nonumber\\
& & \sim A_{0,0}\left[(\phi-1) - 2\omega(\phi - \omega^{-1}) + \omega^2(\phi - \omega^2)\right] + \frac{A_{0,1}}{\rho^6}\left[(\phi - 1)^2 - 2\omega(\phi - \omega^{-1})^2 + \omega^2(\phi - \omega^{-2})^2\right],
\nonumber\\
& &
\end{eqnarray}
where $A_{0,0}=-\frac{\sqrt{3}}{16\pi^4i}\Gamma(\frac{1}{6})\Gamma(\frac{5}{6})$ and $A_{0,1}=\frac{\sqrt{3}}{16\pi^4i}\Gamma(\frac{7}{6})\Gamma(\frac{11}{6})$.

\begin{eqnarray}
\label{eq:confl25}
& & (ii) \varpi_1\sim
\int_{\Gamma^\prime}\frac{d\mu}{8\pi^3}(\Gamma(-\mu))^2\Gamma(\mu+\frac{1}{6})\Gamma(\mu+\frac{5}{6})
\rho^{-6\mu}\nonumber\\
& & \times\frac{\sqrt{3}}{2\pi(\mu+1)}\biggl[2i sin(\pi\mu)\left(e^{-2i\pi\mu}(\phi - 1)^{\mu+1} + \omega(\phi - \omega^{-1})^{\mu+1} - 2\omega^2(\phi - \omega^{-2})^{\mu+1} + (\phi - 1)^{\mu+1}\right)\nonumber\\
& &  + e^{-3i\pi\mu}(\phi - 1)^{\mu + 1}\biggr]
\nonumber\\
& & \sim A_{0,1}\left[(\phi - 1)(A_0 + ln(\rho^{-6})) + i\pi((\phi - 1) + 2\omega(\phi - \omega^{-1} - 4\omega^2(\phi - \omega^{-2}) + (\phi - 1)ln(\phi - 1)\right]\nonumber\\
& & + \frac{A_{0,1}}{\rho^6}\left[(\phi - 1)^2(A_1 + ln(\rho^{-6}) + i\pi\left((\phi - 1) + 2\omega(\phi - \omega^{-1}) - 4\omega^2(\phi - \omega^{-2})\right) + (\phi - 1)^2ln(\phi - 1)\right],\nonumber\\
& &
\end{eqnarray}
where $A_0\equiv -1 - 2\Psi(1) + \Psi(\frac{1}{6}) + \Psi(\frac{5}{6})$ and $A_1\equiv -\frac{1}{2} - 2\Psi(1) - 2 + \Psi(\frac{7}{6})+ \psi(\frac{11}{6})$.

\begin{eqnarray}
\label{eq:confl26}
& & (iii)\ \varpi_2\sim-\int_{\Gamma^\prime}\frac{d\mu}{8\pi^3}(\Gamma(-\mu))^2\Gamma(\mu+\frac{1}{6})\Gamma(\mu+\frac{5}{6})
\rho^{-6\mu}\nonumber\\
& & \times\frac{\sqrt{3}}{2\pi(\mu+1)}\biggl[2i sin(\pi\mu)\left(-2e^{-2i\pi\mu}(\phi - 1)^{\mu + 1} + e^{-2i\pi\mu}\omega(\phi - \omega^{-1})^{\mu+1} + \omega^2(\phi - \omega^{-2})^{\mu+1}\right)\nonumber\\
 & & - \left((\phi - 1)^{\mu+1} - \omega(\phi - \omega^{-1})^{\mu+1}\right) - (\phi - 1)^{\mu+1}\biggr]\nonumber\\
& &
\sim A_{0,0}\biggl[\left(\omega(\phi - \omega^{-1}) - 2(\phi - 1)\right)(A_0 + ln(\rho^{-6})) + 2i\pi\left(-2(\phi - 1) + \omega(\phi - \omega^{-1}) + \omega^2(\phi - \omega^{-2})\right)\nonumber\\
 & & - 2(\phi - 1)ln(\phi - 1)\biggr]
+\frac{A_{0,1}}{\rho^6}\biggl[\left(\omega(\phi - \omega^{-1}) - 2(\phi - 1)^2\right)(A_1 + ln(\rho^{-6}))\nonumber\\
& &  -2i\pi\biggl(-2(\phi - 1)^2 + \omega(\phi - \omega^{-1})^2
+ \omega^2(\phi - \omega^{-2})^2\biggr) - 2(\phi - 1)^2 ln(\phi - 1)\biggr]
\end{eqnarray}

\begin{eqnarray}
\label{eq:confl27}
& & (iv)\ \varpi_3\sim
\int_{\Gamma^\prime}\frac{d\mu}{8\pi^3}(\Gamma(-\mu))^2\Gamma(\mu+\frac{1}{6})\Gamma(\mu + \frac{5}{6})\rho^{-6\mu}e^{i\pi\mu}\nonumber\\
& & \times-\frac{\sqrt{3}}{2\pi(\mu+1)}\left[(\phi - 1)^{\mu+1} - 2\omega(\phi - \omega^{-1})^{\mu+1} + \omega^2(\phi - \omega^{-2})^{\mu+1}\right]\nonumber\\
& & \sim A_{0,0}\biggl[\left((\phi - 1) - 2\omega(\phi - \omega^{-1}) + \omega^2(\phi - \omega^{-2})\right)(A_0 + i\pi + ln(\rho^{-6}))\nonumber\\
& & +(\phi - 1)ln(\phi - 1) - 2\omega(\phi - \omega^{-1})ln(\phi - \omega^{-1}) + \omega^2(\phi - \omega^{-2})ln(\phi - \omega^{-2})\biggr] \nonumber\\
& & + \frac{A_{0,1}}{\rho^6}\biggl[\left((\phi - 1)^2
- 2\omega(\phi - \omega^{-1})^2 + \omega^2(\phi - \omega^{-2})^2\right)(A_1 + i\pi + ln(\rho^{-6}))\nonumber\\
& & +(\phi - 1)^2ln(\phi - 1) - 2\omega(\phi - \omega^{-1})^2ln(\phi - \omega^{-1}) + \omega^2(\phi - \omega^{-2})^2ln(\phi - \omega^{-2})\biggr]
\end{eqnarray}

\begin{eqnarray}
\label{eq:confl28}
& & (v)\ \varpi_4\sim
\int_{\Gamma^\prime}\frac{d\mu}{8\pi^3}(\Gamma(-\mu))^2\Gamma(\mu+\frac{1}{6})\Gamma(\mu + \frac{5}{6})\rho^{-6\mu}e^{i\pi\mu}\nonumber\\
& & \times-\frac{\sqrt{3}}{2\pi(\mu+1)}\left[e^{-2i\pi\mu}(\phi - 1)^{\mu+1} + \omega\omega(\phi - \omega^{-1})^{\mu+1} -2\omega^2(\phi - \omega^{-2})^{\mu+1}\right]\nonumber\\
& & \sim A_{0,0}\biggl[\left((\phi - 1) - 2\omega(\phi - \omega^{-1}) + \omega^2(\phi - \omega^{-2})\right)(A_0 + i\pi + ln(\rho^{-6}))\nonumber\\
& & -2i\pi(\phi - 1) +(\phi - 1)ln(\phi - 1) + \omega(\phi - \omega^{-1})ln(\phi - \omega^{-1})  -2\omega^2(\phi - \omega^{-2})ln(\phi - \omega^{-2})\biggr] \nonumber\\
& & + \frac{A_{0,1}}{\rho^6}\biggl[\left((\phi - 1)^2
- 2\omega(\phi - \omega^{-1})^2 + \omega^2(\phi - \omega^{-2})^2\right)(A_1 + i\pi + ln(\rho^{-6}))\nonumber\\
& & -2i\pi(\phi - 1)^2 +(\phi - 1)^2ln(\phi - 1) + \omega(\phi - \omega^{-1})^2ln(\phi - \omega^{-1}) -  2 \omega^2(\phi - \omega^{-2})^2ln(\phi - \omega^{-2})\biggr]\nonumber\\
& &
\end{eqnarray}

\begin{eqnarray}
\label{eq:confl29}
& & (vi)\ \varpi_5\sim
\int_{\Gamma^\prime}\frac{d\mu}{8\pi^3}(\Gamma(-\mu))^2\Gamma(\mu+\frac{1}{6})\Gamma(\mu + \frac{5}{6})\rho^{-6\mu}e^{i\pi\mu}\nonumber\\
& & \times-\frac{\sqrt{3}}{2\pi(\mu+1)}\left[-2e^{-2i\pi\mu}(\phi - 1)^{\mu+1} + e^{-2i\pi\mu}\omega\omega(\phi - \omega^{-1})^{\mu+1} + \omega^2(\phi - \omega^{-2})^{\mu+1}\right]\nonumber\\
& & \sim A_{0,0}\biggl[\left(-2(\phi - 1) + \omega(\phi - \omega^{-1}) + \omega^2(\phi - \omega^{-2})\right)(A_0 + i\pi + ln(\rho^{-6}))\nonumber\\
& & +4i\pi(\phi - 1) - 2(\phi - 1)ln(\phi - 1) -2i\pi\omega(\phi - \omega^{-1}) + \omega(\phi - \omega^{-1})ln(\phi - \omega^{-1})  + \omega^2(\phi - \omega^{-2})ln(\phi - \omega^{-2})\biggr] \nonumber\\
& & + \frac{A_{0,1}}{\rho^6}\biggl[\left(-2(\phi - 1)^2 +
\omega(\phi - \omega^{-1})^2 + \omega^2(\phi - \omega^{-2})^2\right)(A_1 + i\pi + ln(\rho^{-6}))\nonumber\\
& & 4i\pi(\phi - 1)^2 -2(\phi - 1)^2ln(\phi - 1) - 2i\pi\omega(\phi - \omega^{-1})^2 + \omega(\phi - \omega^{-1})^2ln(\phi - \omega^{-1})\nonumber\\
 & & + \omega^2(\phi - \omega^{-2})^2ln(\phi - \omega^{-2})\biggr].
\end{eqnarray}

\section{Complex Structure Superpotential Extremization}
\setcounter{equation}{0}
\seceqbb

In this appendix, the details pertaining to evaluation of the covariant derivative of the complex structure superpotential in (\ref{eq:Wconfl161}), are given. 

One can see from (\ref{eq:Wconfl12}):
\begin{equation}
\label{eq:Wconfl13}
\hskip - 3cm
\partial_x K\sim \frac{- ln x\left({\bar A_1} B_{42} - {\bar B_{12}} {\bar A_4} + {\bar A_2}B_{52} -
{\bar A_5}B_{22}\right)}{{\cal K}},
\end{equation}
where
\begin{eqnarray}
\label{eq:Wconfl14}
& & {\cal K}\equiv2i\ {\rm Im}\Biggl[ ({\bar A_0} A_3 + {\bar A_1} A_4 + {\bar A}_2 A_5) + ({\bar B_{01}} A_3 + {\bar A_0}B_{31} + {\bar A_1}B_{41} + {\bar B_{11}} A_4 + {\bar A_2} B_{51} + {\bar B_{21}} A_5)x
 \nonumber\\
& &  + ({\bar A_1} B_{42} + {\bar B_{12}} A_4 + {\bar A_2} B_{52} + {\bar B_{22}} A_5)x ln x + ({\bar A_0} C_3 + {\bar C_0} A_3 + {\bar A_1} C_4 + {\bar C_1} A_4 + {\bar A_2} C_5 + {\bar C_2} A_5)(\rho - \rho_0)\Biggr].\nonumber\\
& &
\end{eqnarray}
At the extremum values of the complex structure moduli ($x,\rho - \rho_0$),
\begin{eqnarray}
\label{eq:Wconfl15}
& & \tau=\frac{f^T.{\bar\Pi}}{h^T.{\bar\Pi}}\approx\nonumber\\
& & \frac{1}{(\sum_{i=0}^5h_i{\bar A_i})}\Biggl[f_0\left({\bar A_0} + {\bar B_{01}}{\bar x} + {\bar C_0}({\bar\rho} - {\bar \rho_0})\right) + f_1\left({\bar A_1} + {\bar B_{11}}{\bar x} + {\bar B_{12}}{\bar x} ln{\bar x} + {\bar C_1}({\bar\rho} - {\bar \rho_0})\right)\nonumber\\
 & & + f_2\left({\bar A_2} + {\bar B_{21}}{\bar x} + {\bar B_{22}}{\bar x} ln{\bar x} + {\bar C_2}({\bar\rho} - {\bar \rho_0})\right)+ f_3\left(({\bar A_3} + {\bar B_{31}}{\bar x} + {\bar C_3}({\bar\rho} - {\bar \rho_0})\right)\nonumber\\
& & + f_4\left({\bar A_4} + {\bar B_{41}}{\bar x} + {\bar B_{42}}{\bar x} ln{\bar x} + {\bar C_4}({\bar\rho} - {\bar \rho_0})\right) + f_5\left({\bar A_5} + {\bar B_{51}}{\bar x} + {\bar B_{52}}{\bar x} ln{\bar x} + {\bar C_1}({\bar\rho} - {\bar \rho_0})\right)\Biggr]\nonumber\\
& & \times\Biggl( 1 - \frac{1}{(\sum_{i=0}^5h_i{\bar A_i})}\Biggl[h_0({\bar B_{01}}{\bar x} + {\bar C_0}({\bar\rho} - {\bar \rho_0})) + h_1({\bar B_{11}}{\bar x} + {\bar B_{12}}{\bar x} ln{\bar x} + {\bar C_1}({\bar\rho} - {\bar \rho_0}))\nonumber\\
 & & + h_2({\bar B_{21}}{\bar x} + {\bar B_{22}}{\bar x} ln{\bar x} + {\bar C_2}({\bar\rho} - {\bar \rho_0})) + h_3({\bar A_3} + {\bar B_{31}}{\bar x} + {\bar C_3}({\bar\rho} - {\bar \rho_0}))\nonumber\\
 & & + h_4({\bar A_4} + {\bar B_{41}}{\bar x} + {\bar B_{42}}{\bar x} ln{\bar x} + {\bar C_4}({\bar\rho} - {\bar \rho_0})) + h_5({\bar A_5} + {\bar B_{51}}{\bar x} + {\bar B_{52}}{\bar x} ln{\bar x} + {\bar C_1}({\bar\rho} - {\bar \rho_0}))\Biggr).\nonumber\\
& &
\end{eqnarray}
Hence,
\begin{eqnarray}
\label{eq:Wconfl16}
& & \partial_x W_{c.s.} \approx ln x\Biggl[ {\bar B_{12}}\Biggl(f_1 - \frac{h_1\Xi[f_i;{\bar x},({\bar\rho} - {\bar \rho_0})]}{\sum_{i=0}^5h_i{\bar A_i}} + \frac{h_1(\sum_{j=0}^5f_i{\bar A_i})\Xi[h_i;{\bar x},({\bar\rho} - {\bar \rho_0})]}{(\sum_{i=0}^5h_i{\bar A_i})^2}\Biggr)\nonumber\\
& & + {\bar B_{22}}\Biggl(f_2 - \frac{h_2\Xi[f_i;{\bar x},({\bar\rho} - {\bar \rho_0})]}{\sum_{i=0}^5h_i{\bar A_i}} + \frac{h_2(\sum_{j=0}^5f_i{\bar A_i})\Xi[h_i;{\bar x},({\bar\rho} - {\bar \rho_0})]}{(\sum_{i=0}^5h_i{\bar A_i})^2}\Biggr)\nonumber\\
& & + {\bar B_{42}}\Biggl(f_4 - \frac{h_2\Xi[f_i;{\bar x},({\bar\rho} - {\bar \rho_0})]}{\sum_{i=0}^5h_i{\bar A_i}} + \frac{h_2(\sum_{j=0}^5f_i{\bar A_i})\Xi[h_i;{\bar x},({\bar\rho} - {\bar \rho_0})]}{(\sum_{i=0}^5h_i{\bar A_i})^2}\Biggr)\nonumber\\
& & + {\bar B_{52}}\Biggl(f_5 - \frac{h_5\Xi[f_i;{\bar x},({\bar\rho} - {\bar \rho_0})]}{\sum_{i=0}^5h_i{\bar A_i}} + \frac{h_5(\sum_{j=0}^5f_i{\bar A_i})\Xi[h_i;{\bar x},({\bar\rho} - {\bar \rho_0})]}{(\sum_{i=0}^5h_i{\bar A_i})^2}\Biggr)\Biggr],
\end{eqnarray}
where
\begin{eqnarray}
\label{eq:Xisdefs}
& &
\Xi[f_i; {\bar x},({\bar\rho} - {\bar \rho_0})]\equiv f_0({\bar A_0} + {\bar B_{01}}{\bar x} + {\bar C_0}({\bar\rho} - {\bar \rho_0})) + f_1({\bar A_1} + {\bar B_{11}}{\bar x} + {\bar B_{12}}{\bar x} ln{\bar x} + {\bar C_1}({\bar\rho} - {\bar \rho_0}))\nonumber\\
 & & + f_2({\bar A_2} + {\bar B_{21}}{\bar x} + {\bar B_{22}}{\bar x} ln{\bar x} +
 {\bar C_2}({\bar\rho}-{\bar \rho_0})
 + f_3({\bar A_3}+{\bar B_{31}}{\bar x}+{\bar C_3}({\bar\rho}-{\bar \rho_0}))\nonumber\\
& & + f_4({\bar A_4}+{\bar B_{41}}{\bar x}+{\bar B_{42}}{\bar x} ln{\bar x}+{\bar C_4}({\bar\rho}- {\bar \rho_0}))+f_5({\bar A_5}+{\bar B_{51}}{\bar x}+{\bar B_{52}}{\bar x} ln{\bar x}+{\bar C_1}({\bar\rho}-{\bar \rho_0})),\nonumber\\
& & \equiv f^T{\bar\Pi}({\bar x},{\bar\rho} - {\bar\rho_0})\nonumber\\
& &
\Xi[h_i;{\bar x},({\bar\rho} - {\bar \rho_0})]= h_0({\bar B_{01}}{\bar x} + {\bar C_0}({\bar\rho} - {\bar \rho_0})) + h_1({\bar A_1} + {\bar B_{11}}{\bar x} + {\bar B_{12}}{\bar x} ln{\bar x} + {\bar C_1}({\bar\rho} - {\bar \rho_0}))\nonumber\\
 & & + h_2({\bar B_{21}}{\bar x} + {\bar B_{22}}{\bar x} ln{\bar x} + {\bar C_2}({\bar\rho} - {\bar \rho_0})) + h_3({\bar B_{31}}{\bar x} + {\bar C_3}({\bar\rho} - {\bar \rho_0}))\nonumber\\
& & + h_4({\bar B_{41}}{\bar x} + {\bar B_{42}}{\bar x} ln{\bar x} + {\bar C_4}({\bar\rho} - {\bar \rho_0})) + h_5({\bar A_5} + {\bar B_{51}}{\bar x} + {\bar B_{52}}{\bar x} ln{\bar x} + {\bar C_1}({\bar\rho} - {\bar \rho_0}))\nonumber\\
& & \equiv h^T\left[{\bar\Pi}({\bar x},{\bar\rho} - {\bar\rho_0}) - {\bar\Pi}(x=0,\rho=\rho_0)\right].
\end{eqnarray}

\section{Inverse Metric Components}
\setcounter{equation}{0}
\seceqcc

The components of the inverse of the metric (\ref{eq:nonpert13}), relevant to almost all equations in section {\bf 4} starting from (\ref{eq:nonpert15}) are given as under:

\begin{eqnarray*}
& & (G^{-1})^{\rho_1{\bar\rho_1}}=\nonumber\\
& & \frac{1}{\Delta}\Biggl[
144\,{\cal Y}^2\,{\sqrt{-{\rho_1} + {\bar\rho_1}}}\,
  \biggl( 2\,{\rho_2}\,{\cal X}^2\,{\sqrt{-{\rho_2} + {\bar\rho_2}}} \nonumber\\
  & & -
    \left( 2\,{\cal X}^2 + e^{3\,\phi}\,{\cal X}_1\,{\cal Y}^2 \right) \,{\bar\rho_2}\,
     {\sqrt{-{\rho_2} + {\bar\rho_2}}}
    e^{3\,\phi}\, + {\cal X}_1\,{\cal Y}^2\,\left( 3\,{\sqrt{2}}\,{\cal Y} +
       {\rho_2}\,{\sqrt{-{\rho_2} + {\bar\rho_2}}} \right)  \biggr)\Biggr],\nonumber\\
& & (G^{-1})^{\rho_1{\bar\rho_2}}=\frac{1}{\Delta}\Biggl[
144\,{\cal Y}^2\,\left( -2\,{\cal X}^2 + e^{3\,\phi}\,{\cal X}_1\,{\cal Y}^2 \right) \,
  \left( {\rho_1} - {\bar\rho_1} \right) \,
  \left( {\rho_2} - {\bar\rho_2} \right)\Biggr],\nonumber\\
& & (G^{-1})^{\rho_1{\bar G^1}}=\frac{1}{\Delta}
24\,i  \,e^{\frac{3\,\phi}{2}}\,{\cal X}\,{\cal Y}^2\,\left( {\rho_1} - {\bar\rho_1} \right) \,
  \left( 3\,{\cal Y} + {\sqrt{2}}\,{\rho_2}\,{\sqrt{-{\rho_2} + {\bar\rho_2}}} -
    {\sqrt{2}}\,{\bar\rho_2}\,{\sqrt{-{\rho_2} + {\bar\rho_2}}}
    \right)\nonumber\\
& & (G^{-1})^{\rho_2{\bar\rho_2}}=\nonumber\\
& & \frac{1}{\Delta}
144\,{\cal Y}^2\,\biggl[ -2\,{\rho_1}\,{\cal X}^2\,{\sqrt{-{\rho_1} + {\bar\rho_1}}} +
    \left( 2\,{\cal X}^2 + e^{3\,\phi}\,{\cal X}_1\,{\cal Y}^2 \right) \,{\bar\rho_1}\,
     {\sqrt{-{\rho_1} + {\bar\rho_1}}} \nonumber\\
     & & +
    e^{3\,\phi}\,{\cal X}_1\,{\cal Y}^2\,\left( 3\,{\sqrt{2}}\,{\cal Y} -
       {\rho_1}\,{\sqrt{-{\rho_1} + {\bar\rho_1}}} \right)  \biggr] \,
  {\sqrt{-{\rho_2} + {\bar\rho_2}}},\nonumber\\
& & (G^{-1})^{\rho_2{\bar G^1}}=\frac{1}{\Delta}\Biggl[
-24\,i  \,e^{\frac{3\,\phi}{2}}\,{\cal X}\,{\cal Y}^2\,\left( 3\,{\cal Y} -
    {\sqrt{2}}\,{\rho_1}\,{\sqrt{-{\rho_1} + {\bar\rho_1}}} +
    {\sqrt{2}}\,{\bar\rho_1}\,{\sqrt{-{\rho_1} + {\bar\rho_1}}}
    \right) \,\left( {\rho_2} - {\bar\rho_2} \right)\Biggr],\nonumber\\
& & (G^{-1})^{G^1{\bar G^1}}=\frac{1}{\Delta}\Biggl[
18\,e^{3\,\phi}\,{{k1}}^2\,{\cal X}_1\,{\cal Y}^4 -
  6\,{\sqrt{2}}\,{{k2}}^2\,{\rho_1}\,{\cal X}^2\,{\cal Y}\,
   {\sqrt{-{\rho_1} + {\bar\rho_1}}} -
  3\,{\sqrt{2}}\,e^{3\,\phi}\,{{k1}}^2\,{\rho_1}\,{\cal X}_1\,{\cal Y}^3\,
   {\sqrt{-{\rho_1} + {\bar\rho_1}}} \nonumber\\
   & & +
  6\,{\sqrt{2}}\,{{k_2}}^2\,{\rho_2}\,{\cal X}^2\,{\cal Y}\,
   {\sqrt{-{\rho_2} + {\bar\rho_2}}} +
  3\,{\sqrt{2}}\,e^{3\,\phi}\,{{k_1}}^2\,{\rho_2}\,{\cal X}_1\,{\cal Y}^3\,
   {\sqrt{-{\rho_2} + {\bar\rho_2}}} -
  8\,{{k_2}}^2\,{\rho_1}\,{\rho_2}\,{\cal X}^2\,
   {\sqrt{-{\rho_1} + {\bar\rho_1}}}\,
   {\sqrt{-{\rho_2} + {\bar\rho_2}}}\nonumber\\
   & &  -
  \left( 3\,{\sqrt{2}}\,e^{3\,\phi}\,{{k_1}}^2\,{\cal X}_1\,{\cal Y}^3 +
     2\,{{k_2}}^2\,{\cal X}^2\,\left( 3\,{\sqrt{2}}\,{\cal Y} -
        4\,{\rho_1}\,{\sqrt{-{\rho_1} + {\bar\rho_1}}} \right)  \right) \,
   {\bar\rho_2}\,{\sqrt{-{\rho_2} + {\bar\rho_2}}}\nonumber\\
   & &  +
  {\bar\rho_1}\,{\sqrt{-{\rho_1} + {\bar\rho_1}}}\,
   \left( 3\,{\sqrt{2}}\,e^{3\,\phi}\,{{k_1}}^2\,{\cal X}_1\,{\cal Y}^3 -
     8\,{{k_2}}^2\,{\cal X}^2\,{\bar\rho_2}\,
      {\sqrt{-{\rho_2} + {\bar\rho_2}}} +
     2\,{{k_2}}^2\,{\cal X}^2\,\left( 3\,{\sqrt{2}}\,{\cal Y} +
        4\,{\rho_2}\,{\sqrt{-{\rho_2} + {\bar\rho_2}}} \right)  \right)\Biggr],
        \end{eqnarray*}
with:
\begin{eqnarray*}
& & \Delta=
-18\,e^{3\,\phi}\,{\cal X}_1\,Y^4 + 6\,{\sqrt{2}}\,{\rho_1}\,{\cal X}^2\,{\cal Y}\,
   {\sqrt{-{\rho_1} + {\bar\rho_1}}} +
  3\,{\sqrt{2}}\,e^{3\,\phi}\,{\rho_1}\,{\cal X}_1\,{\cal Y}^3\,
   {\sqrt{-{\rho_1} + {\bar\rho_1}}}\nonumber\\
   & &  -
  6\,{\sqrt{2}}\,{\rho_2}\,{\cal X}^2\,{\cal Y}\,{\sqrt{-{\rho_2} + {\bar\rho_2}}} -
  3\,{\sqrt{2}}\,e^{3\,\phi}\,{\rho_2}\,{\cal X}_1\,{\cal Y}^3\,
   {\sqrt{-{\rho_2} + {\bar\rho_2}}} +
  8\,{\rho_1}\,{\rho_2}\,X^2\,{\sqrt{-{\rho_1} + {\bar\rho_1}}}\,
   {\sqrt{-{\rho_2} + {\bar\rho_2}}} \nonumber\\
   & & +
  \left( 3\,{\sqrt{2}}\,e^{3\,\phi}\,{\cal X}_1\,{\cal Y}^3 +
     {\cal X}^2\,\left( 6\,{\sqrt{2}}\,{\cal Y} - 8\,{\rho_1}\,{\sqrt{-{\rho_1} + {\bar\rho_1}}}
        \right)  \right) \,{\bar\rho_2}\,
   {\sqrt{-{\rho_2} + {\bar\rho_2}}} \nonumber\\
   & & -
  {\bar\rho_1}\,{\sqrt{-{\rho_1} + {\bar\rho_1}}}\,
   \left( 3\,{\sqrt{2}}\,e^{3\,\phi}\,{\cal X}_1\,{\cal Y}^3 -
     8\,{\cal X}^2\,{\bar\rho_2}\,{\sqrt{-{\rho_2} + {\bar\rho_2}}} +
     {{\cal X}}^2\,\left( 6\,{\sqrt{2}}\,{\cal Y} + 8\,{\rho_2}\,
         {\sqrt{-{\rho_2} + {\bar\rho_2}}} \right)  \right);\nonumber\\
         & & {\cal X}\equiv \sum_c\sum_{(n,m)\in{\bf Z}^2/(0,0)}A_{n,m,n_{k^c}}(\tau)sin(nk.b + mk.c).
         \end{eqnarray*}

\section{Ingredients for Evaluation of ${\cal N}_{IJ}$}
\setcounter{equation}{0}
\seceqdd

In this appendix we fill in the details relevant to evaluation of $X^I Im(F_{IJ}) X^J$ to arrive at (\ref{eq:BHinv15}).

First, using (\ref{eq:BHinv12}), one arrives at;
\begin{eqnarray}
\label{eq:BHinv13}
& &  Im(F_{0I})X^I=-\frac{i}{2}\Biggl[\left(\frac{B_{01}}{B_3}+\frac{C_0}{C_3}-\frac{{\bar B_{01}}}{{\bar B_3}}-\frac{{\bar C_0}}{{\bar C_3}}\right)(A_3+B_{31}x+C_3(\rho-\rho_0)\nonumber\\
& & +\left(\frac{C_1}{C_3}-\frac{{\bar C_1}}{{\bar C_3}}\right)(A_4+B_{41}x+B_{42}x ln x+C_4(\rho-\rho_0))\Biggr] +\left(\frac{C_2}{C_3}-\frac{{\bar C_2}}{{\bar C_3}}\right)(A_5+B_{51}x+B_{52}x ln x+C_5(\rho-\rho_0))\Biggr];\nonumber\\
& & Im(F_{1I})X^I=-\frac{i}{2}\Biggl[\left(\frac{C_1}{C_3}-\frac{{\bar C_1}}{{\bar C_3}}\right)(A_3+B_{31}x+C_3(\rho-\rho_0)\nonumber\\
& & +\left(\frac{C_1}{C_4}-\frac{{\bar C_1}}{{\bar C_4}}\right)(A_4+B_{41}x+B_{42}x ln x+C_4(\rho-\rho_0))\Biggr] +\left(\frac{C_2}{C_4}-\frac{{\bar C_2}}{{\bar C_4}}\right)(A_5+B_{51}x+B_{52}x ln x+C_5(\rho-\rho_0))\Biggr];\nonumber\\
& & Im(F_{2I})X^I=-\frac{i}{2}\Biggl[\left(\frac{C_2}{C_3}-\frac{{\bar C_2}}{{\bar C_3}}\right)(A_3+B_{31}x+C_3(\rho-\rho_0)\nonumber\\
& & +\left(\frac{C_2}{C_4}-\frac{{\bar C_2}}{{\bar C_4}}\right)(A_4+B_{41}x+B_{42}x ln x+C_4(\rho-\rho_0))\Biggr] +\left(\frac{C_2}{C_5}-\frac{{\bar C_2}}{{\bar C_5}}\right)(A_5+B_{51}x+B_{52}x ln x+C_5(\rho-\rho_0))\Biggr].\nonumber\\
& &
\end{eqnarray}
This hence yields
\begin{eqnarray}
\label{eq:BHinv14}
& & X^I Im(F_{IJ} X^J = (X^0)^2 Im(F_{00}) + (X^1)^2 Im(F_{11}) + (X^2)^2 Im(F_{22})\nonumber\\
& & + 2x^0X1 Im(F_{01}) + 2X^0X^2 Im(F_{02}) + 2X^1X^2 Im(F_{12})\nonumber\\
& & \approx-\frac{i}{2}\Biggl[\Biggl(A_3^2 + 2A_3B_{31} x + 2A_3C_3 (\rho-\rho_0)\Biggr)\left(\frac{B_{01}}{B_3} + \frac{C_0}{C_3} - \frac{{\bar B_{01}}}{{\bar B_3}} - \frac{{\bar C_0}}{{\bar C_3}}\right)\nonumber\\
& & + \Biggl(A_4^2 + 2A_4B_{41} x + 2A_4B_{42} x ln x + 2A_4C_4 (\rho-\rho_0)\Biggr)\left(\frac{C_1}{C_4} - \frac{{\bar C_1}}{{\bar C_4}}\right)\nonumber\\
& & + \Biggl(A_5^2 + 2A_5B_{51} x + 2A_5B_{52} x ln x + 2A_5C_5 (\rho-\rho_0)\Biggr)
\left(\frac{C_2}{C_4} - \frac{{\bar C_2}}{{\bar C_5}}\right)\nonumber\\
& & + \Biggl(A_3A_4 + [A_3B_{41}+A_4B_{31}] x + A_3B_{42} x ln x + [A_3C_4 + A_4C_3](\rho - \rho_0)\Biggr)\left(\frac{C_1}{C_3} - \frac{{\bar C_1}}{{\bar C_3}}\right)\nonumber\\
& & + \Biggl(A_3A_5 + [A_3B_{51}+A_5B_{31}] x + A_3B_{52} x ln x + [A_3C_5 + A_5C_3](\rho - \rho_0)\Biggr)\left(\frac{C_2}{C_3} - \frac{{\bar C_2}}{{\bar C_3}}\right)\nonumber\\
& & + \Biggl(A_4A_5 + [A_4B_{51}+A_5B_{41}] x + A_4B_{52} x ln x + [A_4C_5 + A_5C_4](\rho - \rho_0)\Biggr)\left(\frac{C_2}{C_4} - \frac{{\bar C_2}}{{\bar C_4}}\right)\Biggr].
\end{eqnarray}

\newpage
\begin{table}[htbp]
\begin{tabular}{|c|c|} \hline
$(G^{-1})^{\rho_1{\bar\rho}_1}|\partial_{\rho_1}W_{np}|^2$ &
$\frac{{\cal V}\sqrt{ln {\cal V}}}{{\cal V}^{2n^1e^{-\phi}}}$ \\ \hline
$(G^{-1})^{\rho_2{\bar\rho}_2}|\partial_{\rho_2}W_{np}|^2$ &
${\cal V}^{\frac{4}{3}}e^{-2n^2e^{-\phi}{\cal V}^{\frac{2}{3}}}$ \\ \hline
$(G^{-1})^{\rho_1{\bar\rho_2}}\partial_{\rho_1}W_{np}{\bar\partial}_{\bar\rho_2}{\bar W}_{np} + c.c.$
&
$\frac{{\cal V}^{\frac{2}{3}} ln {\cal V} e^{-n^2e^{-\phi}{\cal V}^{\frac{2}{3}}}}{{\cal V}^{n^1e^{-\phi}}}$
\\ \hline
$(G^{-1})^{G^1{\bar G^1}}|\partial_{G^1}W|^2 + (G^{-1})^{G^2{\bar G^2}}|\partial_{G^2}W|^2 + $ &
${\cal V}^{1-(2n^1\ or\ 2n^2\ or\ n^1+n^2)e^{-\phi}}$\\
$(G^{-1})^{G^1{\bar G^2}}\partial_{G^1}W_{np}{\bar\partial}_{\bar G^2}{\bar W}_{np} + c.c.$ & \\ \hline
$(G^{-1})^{\rho_1{\bar G^1}}\partial_{\rho_1}W_{np}{\bar\partial}_{\bar G^1}{\bar W}_{np} + c.c.$
& $\frac{ln {\cal V} }{{\cal V}^{2n^1e^{-\phi}}}$ \\ \hline
$(G^{-1})^{\rho_2{\bar G^1}}\partial_{\rho_2}W_{np}{\bar\partial}_{\bar G^1}{\bar W}_{np} + c.c.$
& $\frac{{\cal V}^{\frac{2}{3}} e^{-n^1e^{-\phi}{\cal V}^{\frac{2}{3}}}}{{\cal V}^{n^1e^{-\phi}+\frac{1}{3}}}$ \\ \hline
\end{tabular}
\caption{$(G^{-1})^{A{\bar B}}\partial_A W_{np}{\bar\partial}_{\bar B}{\bar W}_{np}$}
\end{table}

\begin{table}[htbp]
\begin{tabular}{|c|c|} \hline
$(G^{-1})^{\rho_1{\bar\rho}_1}\partial_{\rho_1}K{\bar\partial}_{\bar\rho_1}{\bar W}_{np}+c.c.$ &
$\frac{ln {\cal V}}{{\cal V}^{n^1e^{-\phi}}}$ \\ \hline
$(G^{-1})^{\rho_2{\bar\rho}_2}\partial_{\rho_2}K{\bar\partial}_{\bar\rho_1}{\bar W}_{np}+c.c.$ &
${\cal V}^{\frac{2}{3}}e^{-n^2e^{-\phi}{\cal V}^{\frac{2}{3}}}$ \\ \hline
$(G^{-1})^{\rho_1{\bar\rho}_2}\partial_{\rho_1}K{\bar\partial}_{\bar\rho_2}{\bar W}_{np}+c.c.$ &
$\frac{(ln {\cal V})^{\frac{3}{2}}e^{-n^2e^{-\phi}{\cal V}^{\frac{2}{3}}}}{{\cal V}^{\frac{1}{3}}}$
\\ \hline
$(G^{-1})^{G^1{\bar G^1}}\partial_{G^1}K{\bar\partial}_{\bar G^1}{\bar W}_{np} +
(G^{-1})^{G^1{\bar G^1}}\partial_{G^1}K{\bar\partial}_{\bar G^1}{\bar W}_{np} +$ & $\frac{1}{{\cal V}^{n^1e^{-\phi}}}$
\\
$(G^{-1})^{G^1{\bar G^2}}\partial_{G^1}K{\bar\partial}_{\bar G^2}{\bar W}_{np} +
(G^{-1})^{\rho_2{\bar G^1}}\partial_{\rho_2}K{\bar\partial}_{\bar G^1}{\bar W}_{np} + c.c.$
 & \\ \hline
$(G^{-1})^{\rho_1{\bar G^1}}\partial_{\rho_1}K{\bar\partial}_{\bar G^1}{\bar W}_{np} + c.c.$
& $\frac{(ln {\cal V})^{\frac{3}{2}}}{{\cal V}^{1+n^1e^{-\phi}}}$ \\ \hline
\end{tabular}
\caption{$(G^{-1})^{A{\bar B}}\partial_AK{\bar\partial}_{\bar B}{\bar W}_{np} + c.c.$}
\end{table}

\begin{table}[htbp]
\begin{tabular}{|c|c|} \hline
$(G^{-1})^{\rho_1{\bar\rho}_1}|\partial_{\rho_1}K|^2$ & $\frac{(ln {\cal V})^{\frac{3}{2}}}{{\cal V}}$ \\ \hline
$(G^{-1})^{\rho_2{\bar\rho}_2}|\partial_{\rho_2}K|^2$  & ${\cal O}(1)$ \\ \hline
$(G^{-1})^{\rho_1{\bar\rho}_2}\partial_{\rho_1}{\bar\partial}_{\bar\rho_2}K + c.c.$
& $\frac{(ln {\cal V})^{\frac{3}{2}}}{{\cal V}}$ \\ \hline
$(G^{-1})^{G^1{\bar G^1}}|\partial_{G^1}K|^2 + (G^{-1})^{G^2{\bar G^2}}|\partial_{G^2}K|^2 + $ &
$\frac{1}{\cal V}$\\
$(G^{-1})^{G^1{\bar G^2}}\partial_{G^1}K{\bar\partial}_{\bar G^2}K
+ (G^{-1})^{\rho_2{\bar G^1}}\partial_{\rho_2}K{\bar\partial}_{\bar G^1}K + c.c.$
 & \\ \hline
$(G^{-1})^{\rho_1{\bar G^1}}\partial_{\rho_1}K{\bar\partial}_{\bar G^1}K + c.c.$
& $\frac{(ln {\cal V})^{\frac{3}{2}}}{{\cal V}^2}$
\\ \hline
\end{tabular}
\caption{$(G^{-1})^{A{\bar B}}\partial_AK{\bar\partial}_{\bar B}K$}
\end{table}

\end{document}